\documentclass[
 reprint,
 amsmath,amssymb,
 aps,
prb,
]{revtex4-2}

\usepackage{graphicx}
\usepackage{dcolumn}
\usepackage{bm}
\usepackage{xcolor}

\usepackage{soul}

\newcommand{\revise}{}
\usepackage{amsmath}
\begin{document}

\title{The Bulk Penetration of Edge Properties in Two-Dimensional Materials}

\author{Markus Kari}
\affiliation{Nanoscience Center, Department of Physics, University of Jyv\"askyl\"a, Finland.}
\author{Pekka Koskinen}
\email{pekka.j.koskinen@jyu.fi}
\affiliation{Nanoscience Center, Department of Physics, University of Jyv\"askyl\"a, Finland.}

\date{\today}

\begin{abstract}
Edges are essential for the mechanical, chemical, electronic, and magnetic properties of two-dimensional (2D) materials. \revise{Research has shown that features assigned to edges are not strictly localized but often penetrate the bulk to some degree. However, mechanical edge properties, such as edge energies and stresses, are typically assigned at the system level, with spatial bulk penetrations that remain unknown.} Here, we use density-functional tight-binding simulations to study how deep various edge properties spatially penetrate the 2D bulk. We study nine different edges made of four materials: graphene, goldene, boron nitride, and molybdenum disulfide. By investigating edge energies, edge stresses, and edge elastic moduli, we find that although the edge properties typically originate near edges, they still penetrate the bulk to some degree. An utmost example is goldene with a staggered edge, whose edge properties penetrate the bulk nanometer-deep. Our results caution against associating system-level edge properties too strictly with the edge, especially if those properties are further used in continuum models.
\end{abstract}

\maketitle


\section{Introduction}
During the last couple of decades, the properties of many two-dimensional (2D) materials have been found to be critically dependent on their edges \cite{graphene_edge_review, edge_review_newer}. Free edges sustain many important phenomena such as adsorption \cite{adsorption1, adsorption2}, sensing \cite{sensors}, stabilization \cite{Heterostructure_application}, catalysis \cite{catalysis}, chemical reduction \cite{MoS2_synthesis}, and 2D material synthesis \cite{synthesis}. Edges also play a central role in lateral heterostructures \cite{heterostructure}. 

\revise{Given their significant role, the character of edge properties has been investigated both theoretically and experimentally. For example, scanning probe experiments have demonstrated that electronic edge states are not strictly localized, but penetrate the bulk depending on the material and nanostructure geometry \cite{tao2011spatially}. Prominent examples are the zigzag edge states in graphene \cite{fujita1996peculiar,kobayashi2006edge,wang2016giant} and metallic edge states in MoS$_2$ \cite{bollinger2001one,lauritsen2004hydrodesulfurization}. From an energetic and mechanical point of view, the essential edge properties are edge energy, edge stress, and edge elastic modulus \cite{Reddy2009}.} However, these properties are typically assigned at the system level, which may cause certain problems.

\begin{figure}[b!]
    \centering
    \includegraphics[width=\linewidth]{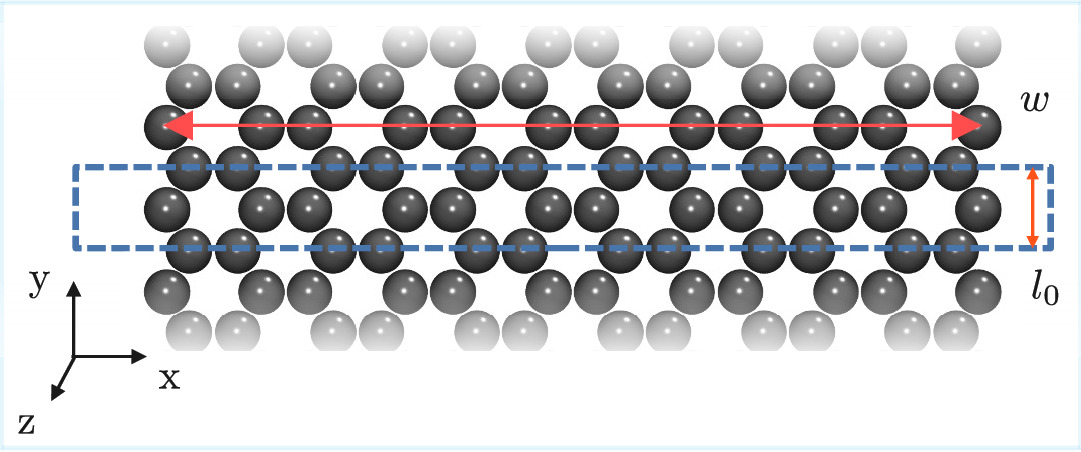}
    \caption{Schematic of a nanoribbon with two edges. The simulation cell is periodic in the $y$-direction, with unstrained length $l_0$ corresponding to 2D bulk bond lengths in the middle. The width $w$ is measured from the outermost atoms. }
    \label{fig:setup}
\end{figure}

To see the nature of these problems, let us start by considering a periodic cell of an infinitely long and very wide nanoribbon (Fig.~\ref{fig:setup}). The cell length $l_0$ derives from the 2D bulk at the ribbon's interior. The ribbon energy per unit length is
\begin{equation}
E_0/l_0=(E_{\text{bulk}}+E_{\text{edge}})/l_0=N\varepsilon_{2D}/l_0+2\lambda,
\label{eq:elastic_C}
\end{equation}
where $N$ is the number of atoms in the cell, $\varepsilon_{2D}$ the 2D bulk binding energy, and $\lambda$ the edge energy, the energy cost per unit length for creating the edge \cite{Gan2010}. The factor of two accounts for the ribbon's two edges. When the ribbon is axially strained by $\varepsilon=(l-l_0)/l_0$, the energy per unit length can be expanded as  
\begin{equation}
E(\varepsilon)/l_0=E_0/l_0+\alpha\varepsilon+\beta\varepsilon^2.
\label{eq:energy_first}
\end{equation}
First, since the bulk of the ribbon must have no stress, the linear term is naturally associated with edges so that $\alpha=2\tau$, where $\tau$ is the edge stress \cite{Huang2009}. Second, the quadratic term refers to ribbons' elastic behavior. The term has both bulk and edge contributions,
\begin{equation}
(\beta_{\text{bulk}}+\beta_{\text{edge}})\varepsilon^2 =(\frac{1}{2}Yw+2\cdot \frac{1}{2}Y_e)\varepsilon^2,
\label{eq:elastic_A}
\end{equation}
where $Y$ is the 2D Young's modulus, $Y_e$ is the (1D) edge elastic modulus, and $w$ is ribbon width, conventionally defined by $w=\max_i \{x_i\}-\min_i \{x_i\}$, with $x_i$ as the $x$-coordinate of atom $i$  \cite{Reddy2009}. 

The energy density of a strained ribbon is thus conventionally written as \cite{abidi2025}
\begin{equation}
    E(\varepsilon)/l_0=N \varepsilon_{2D}/l_0+2\lambda+2\tau\varepsilon+\frac{1}{2}Yw\varepsilon^2+Y_e \varepsilon^2.
\label{eq:Eperl}
\end{equation}
With $\varepsilon_{2D}$ and $Y$ known from the 2D bulk, the edge energies, stresses, and elastic moduli can be fitted to ribbons under varying strains. This fitting process is well established in the literature \cite{Gan2010, Huang2009, Reddy2009, abidi2025, Huang2012, Qi2013, Yang2021}. However, note that Eq.~(\ref{eq:Eperl}) describes the entire system so that the so-called edge properties are defined not for the edges but for the system. In principle, there is nothing in the derivation of the equations that demands the edge properties to be located at the edges. \revise{The spatial characters and the bulk penetration lengths of these mechanical edge properties remain unknown.}

Therefore, in this article, we use electronic structure simulations to address the following question: \revise{\emph{How deep do the mechanical edge properties penetrate the bulk?}} We address the question by studying nine different edges made of four materials: graphene, goldene, boron nitride, and molybdenum disulfide. We also investigate how the electronic structures of ribbons express the presence of their edges. \revise{We find that electronic properties get localized near edges, unless they get affected by long-range Coulomb interactions. While mechanical edge properties usually penetrate the bulk much less than a nanometer, for some edges the penetration length can be much greater}.

\section{Systems and methods}
To represent different 2D materials, we chose four well-known materials: graphene, golden (monolayer hexagonal Au), hexagonal boron nitride (hBN), and molybdenum disulfide (MoS$_2$). They were chosen to represent different material categories: graphene is a semimetal, goldene is a metal, and MoS$_2$ is a semiconductor; graphene, goldene, and hBN are atomically thin and MoS$_2$ has a sandwich structure; goldene has metallic bonding and others are bound primarily covalently; all materials can have different edges; and all the materials have been synthesized \cite{graphene_synthesis, goldene_2024, hBN_synthesis, MoS2_synthesis}.

\begin{figure}[b!]
    \centering
    \includegraphics[width=\linewidth]{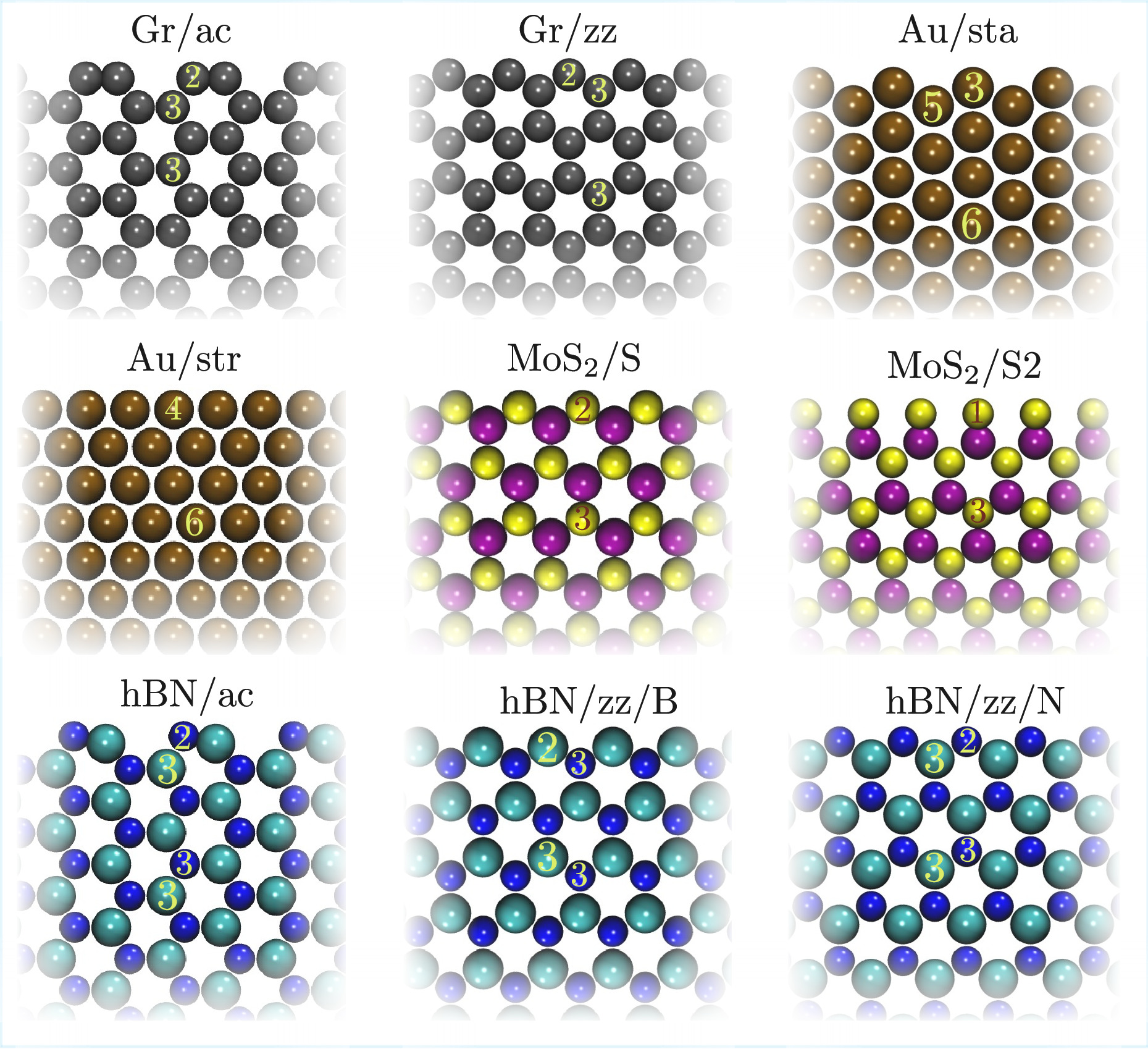}
    \caption{The four materials and nine edges studied: armchair (Gr/ac) and zigzag (Gr/zz) edges of graphene, staggered (Au/sta) and straight (Au/str) edges of goldene, two different S-terminated edges of MoS$_2$ (MoS$_2$/zz/S and MoS$_2$/zz/S2) \cite{Cao2015}, and armchair (hBN/ac), zigzag B-terminated (hBN/zz/B), and zigzag N-terminated (hBN/zz/N) edges of hBN. 
    The numbers show atoms' coordination numbers.}
    \label{fig:systems}
\end{figure}

For each material, we studied two or three different edges. For clarity, we focused on pristine, unpassivated edges. Termination by hydrogen, for example, could alter edge properties significantly \cite{Gan2010}. We aimed for both the most stable and the most studied edge types with the most reference values. The studied edges are shown in Fig.~\ref{fig:systems}.

The edges were simulated using self-consistent-charge tight-binding density-functional theory (SCC-DFTB) \cite{SCC}. DFTB has been shown to describe these four materials well, and its accuracy is adequate for our purposes \cite{koskinen2012graphene,koskinen2007liquid,koskinen2015plenty,koskinen2014density,zobelli2007vacancy}. In particular, as discussed later, DFTB lends itself to an easy and transparent analysis of various properties' spatial dependence \cite{hotbit}. The spatial analysis is critical for addressing our question and would be challenging to accomplish using density-functional theory. At the same time, DFTB still accounts for all the non-local quantum effects, which are essential for addressing our question reliably. Although machine learning potentials are very much in vogue, they and other force fields could not describe quantum effects, charge transfer effects, or any non-local effects, so central to our question \cite{ff, ml}. 

The DFTB simulations were done using the Hotbit code \cite{hotbit}. The ribbons were modeled by the simulation cell shown in Fig.~\ref{fig:systems} with $1\times 20\times 1$ $k$-point sampling and $0.05$~eV Fermi broadening. The BN parametrizations were adopted from the matsci package \cite{matsci, hBN_dftb}, others were native to Hotbit \cite{hotbit, hotbit_param_old, hotbit_param_mos2, hotbit_param_gold}. All systems were relaxed by the BFGS algorithm to forces below $1$~meV/\AA\ \cite{nocedal1999numerical}.

\begin{table}[t!]
\caption{\label{tab:width}
Number of ribbons, the minimum and maximum ribbon widths, and the maximum (absolute) strains used in the calculations of edge properties.}
\begin{ruledtabular}
\begin{tabular}{lcccc}
Edge  & nr.     & $\min w$ (\AA)  & $\max w$ (\AA)  & $\varepsilon_{\textrm{max}}$ (\%) \\ \colrule
Gr/ac       & 25 & 18  & 48      & 0.5             \\
Gr/zz       & 13 & 16  & 41      & 0.5             \\
Au/sta      & 7  & 21  & 51      & 2.0               \\
Au/str      & 8  & 38  & 69      & 1.5             \\
MoS$_2$/S   & 5  & 23  & 34      & 1.0               \\
MoS$_2$/S2 & 5 & 23 & 34      & 1.0              \\
hBN/ac      & 17 & 18  & 39      & 2.0               \\
hBN/zz/B    & 9  & 16  & 34      & 2.0               \\
hBN/zz/N    & 9  & 16  & 34      & 2.0               
\end{tabular}
\end{ruledtabular}
\end{table}

The system-level edge properties were calculated as usual. The edge energies were obtained by calculating unstrained ribbons of different widths and using Eq.~(\ref{eq:elastic_C}) to fit the 2D cohesion energy and edge energy simultaneously, following Ref.~\onlinecite{abidi2025}. The ranges of ribbon widths used in the calculations are shown in Table~\ref{tab:width}.

The system-level edge stresses, edge elastic moduli, and bulk moduli were obtained by calculating the energies of strained ribbons and fitting $\tau$, $Y$, and $Y_e$ to the data ($11$ strains $\varepsilon \in [-\varepsilon_{max},\varepsilon_{max}]$, see Table~\ref{tab:width}). (Because of inevitable numerical inaccuracy in determining bulk $l_0$, a tiny uncertainty of zero strain was allowed in the fit \cite{abidi2025}.) \revise{All the atoms were always fully relaxed for given strains, by the definition of Young's modulus. (Also ribbon widths changed due to the Poisson effect; however, here the values of the Poisson ratios are irrelevant.)} Through Eq.~(\ref{eq:Eperl}), these fits then provided a fully analytical system-level description of energy $E(\varepsilon)$.

At the same time, \revise{DFTB allows expressing the total energy as
\begin{equation}
    E_{DFTB}(\varepsilon)=\sum_{i=1}^N e_i(\varepsilon),
    \label{eq:edftb}
\end{equation}
where $N$ is the number of atoms and $e_i$'s are the local contribution to the total energy by an atom at ${\bf r}_i$. \cite{hotbit}. (See Appendix for a technical discussion of the expression.)}. Juxtaposing Eqs.~(\ref{eq:Eperl}) and (\ref{eq:edftb}), we can decompose the system-level edge properties into spatial contributions from local atomic properties. The edge energy becomes
\begin{equation}
    \lambda=\sum_{i=1}^N \lambda_i(0),
\end{equation}
where $\lambda_i(\varepsilon)= [e_i(\varepsilon)-\varepsilon_{2D}]/l_0$ is the spatial edge energy contribution at ${\bf r}_i$. \revise{The function $\lambda_i(\varepsilon)$ was constructed by a quadratic fit to $11$ different values of strain, which also allowed evaluating the first $\lambda'_i(0)=d\lambda(\varepsilon)/d\varepsilon|_{\varepsilon=0}$ and second derivatives $\lambda''_i(0)=d^2\lambda(\varepsilon)/d\varepsilon^2|_{\varepsilon=0}$.} Accordingly, the edge stress becomes
\begin{equation}
    \tau = \frac{1}{2}\sum_{i=1}^N \lambda'_i(0)
\end{equation}
with spatial stress contributions $\tau_i=\lambda'_i(0)/2$ at ${\bf r}_i$. The two elastic moduli can be expressed as
\begin{equation}
    Yw+2Y_e = \sum_{i=1}^N \lambda''_i(0).
    \label{eq:elastic}
\end{equation}
The two terms can be decoupled by noting that for 2D bulk $Yw_{bulk}=N\lambda''_{bulk}$, where $\lambda''_{bulk}$ is here calculated as the average of $\lambda''_i(0)$ across the middle one-third-ribbon of width $w_{bulk}$. Young's modulus can also be expressed as $Y=\lambda''_{bulk}/w_1$, where $w_1$ is the nominal width of one atom in the ribbon. Consequently, the edge elastic modulus becomes
\begin{equation}
    Y_e=\sum_{i=1}^N \frac{1}{2}\left(\lambda''_i(0)-\lambda''_{bulk}\right)
\end{equation}
with spatial contributions $Y_{e,i}=\frac{1}{2}(\lambda''_i(0)-\lambda''_{bulk})$ at ${\bf r}_i$. Using these equations, the spatial contributions to edge properties were adopted as the averaged contributions from the three widest ribbons (Table~\ref{tab:width}).  

As a final methodological note, the edge energy contributions will be presented at atomic resolution. However, the edge stress and elastic modulus contributions will be presented at atomic resolution only for graphene and goldene and at structural unit resolution for hBN and MoS$_2$. (For example, within the hBN bulk, the edge stress contribution $\lambda'_i(0)$ can be positive for B and negative for N, so they need to be averaged out; edge energy contributions don't have this complexity.)

\section{Results}

We begin by presenting the edge properties calculated \revise{in the conventional} way. Edge energies, edge stresses, Young's moduli, and edge elastic moduli are shown in Table~\ref{tab:data}. The obtained edge properties are in fair agreement with earlier studies, as will be discussed with their spatial contributions in what follows.

\subsection{Edge energy}

\begin{table}
\caption{System-level edge energies ($\lambda$), edge stresses ($\tau)$, Young’s moduli ($Y$), edge elastic moduli ($Y_e$), and effective edge widths ($w_e$) for the studied edges.}
\begin{ruledtabular}
\begin{tabular}{lrrrrr}                                      
           Edge & \begin{tabular}[c]{@{}c@{}} $\lambda$ \\ (eV/Å)\end{tabular} & \begin{tabular}[c]{@{}c@{}} $\tau$ \\ (eV/Å)\end{tabular} & \begin{tabular}[c]{@{}c@{}} $Y$ \\(eV/Å$^2$)\end{tabular} & \begin{tabular}[c]{@{}c@{}} $Y_e$ \\(eV/Å)\end{tabular} & \begin{tabular}[c]{@{}c@{}} $w_e$ \\ (Å) \end{tabular} \\ \hline                  
Gr/ac       & 1.10   & -1.70   & 25.4  & 2.0  &   -1.1 \\
Gr/zz       & 1.30   & -0.55  & 25.4  & 32.9   &    0.5 \\
Au/sta      & 0.25   & 0.11   & 3.8   & -30.6   &   -0.3\\
Au/str      & 0.21   & 0.17   & 4.0   & -1.2     &  1.3 \\
hBN/ac      & 0.87   & -0.62  & 17.8  & -6.9  &   -0.9\\
hBN/zz/B    & 1.31   & -1.00  & 17.8  & 29.9   &    1.1\\
hBN/zz/N    & 1.26   & 0.42   & 17.8  & 11.1    &    0.1\\
MoS$_2$/zz/S  & 0.72   & -1.69  & 8.7   & 16.5    &  0.8  \\
MoS$_2$/zz/S2 & 1.95   & 0.67   & 8.7   & 0.4   & -1.1
\end{tabular}
\end{ruledtabular}
\label{tab:data}
\end{table}

\begin{figure}[b!]
    \centering
    \includegraphics[width=\linewidth]{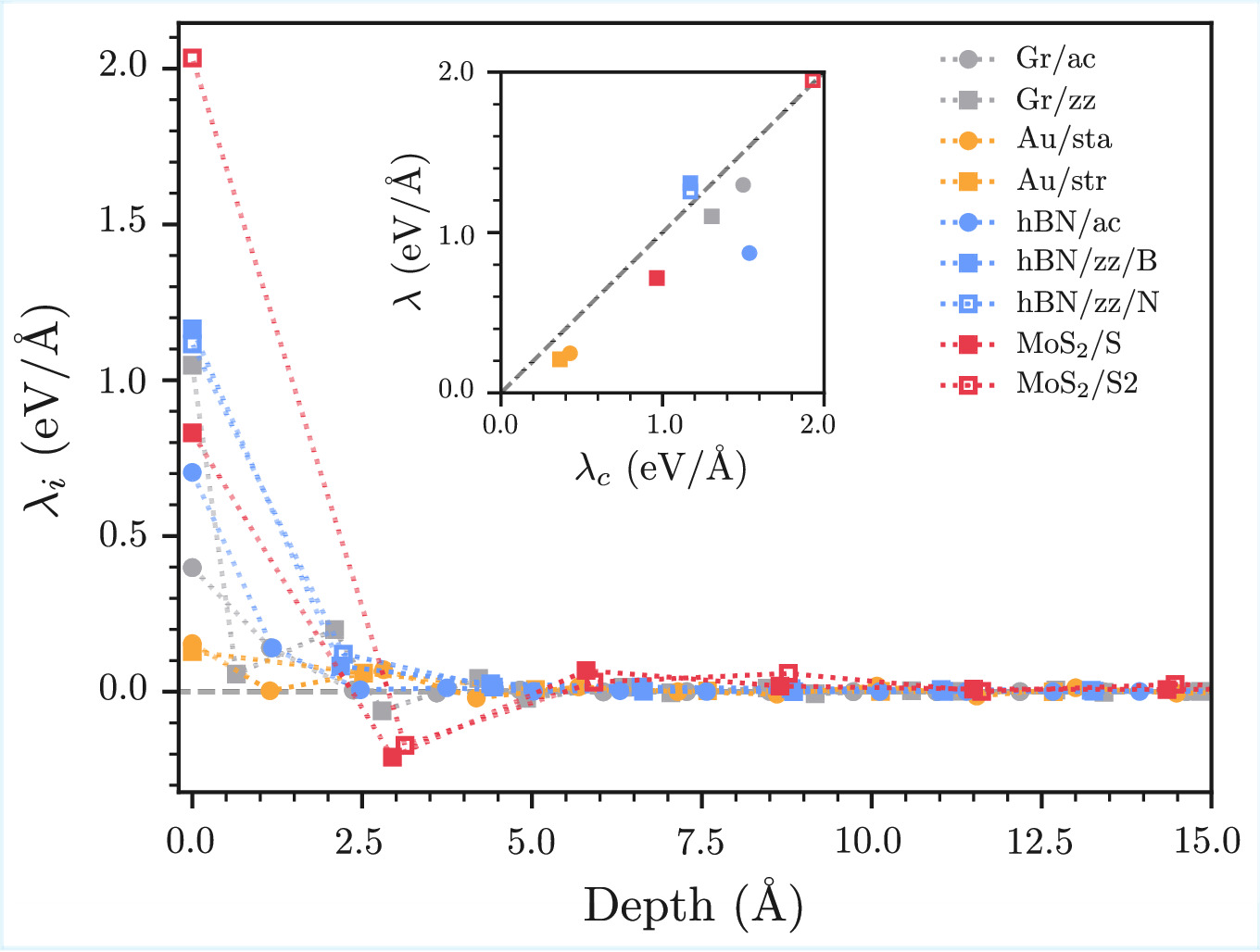}
    \caption{Spatial contributions to edge energies as a function of bulk depth. Inset: The correlation between calculated edge energies ($\lambda$) and estimates based on coordination numbers [$\lambda_C$ from Eq.~(\ref{eq:gamma_model})]. }
    \label{fig:edge_e}
\end{figure}

All edge energies are positive, indicating that edge formations cost energy (Table~\ref{tab:data}). The edge energies of ac and zz edges for graphene and hBN are similar, ac having a bit smaller energy for both materials. \revise{Values are in relative agreement with literature, which report energies between $0.34$ and $1.54$ eV/\AA$^2$ for Gr(zz), between $0.29$ and $1.2$ eV/\AA$^2$ for Gr(ac), between $0.36$ and $1.28$ eV/\AA$^2$ for hBN(zz), and between $0.23$ and $0.76$ eV/\AA$^2$ for hBN(ac) \cite{Huang2009, Gan2010, Yang2021, Huang2012}. Direct comparisons are difficult because edge energies in the literature depend considerably on the computational methods.} The edge energies for goldene are slightly smaller than those from density-functional calculations \cite{abidi2025}. MoS$_2$/zz/S has edge energy close to those of graphene and hBN \cite{Qi2013}, whereas the edge energy of MoS$_2$/zz/S2 is the highest.

The trends in the edge energies with respect to lattices and edge types are reasonably captured by a simple linear bond-cutting model based on atoms' coordination numbers $C_i$ \cite{bondCutting1992calculated, CohCoord2017}. This model approximates the edge energy as
\begin{equation}
    \lambda_C =|\varepsilon_{2D}| \cdot \frac{C_b- C_e}{C_b}\eta,
    \label{eq:gamma_model}
\end{equation}
where $C_b$ is the mean coordination number of the 2D bulk, $C_e$ is the mean coordination number of undercoordinated edge atoms, and $\eta$ is the edge atom density per unit length \cite{Bond-cuttingmclachlan1957surface, bondCutting1992calculated}. Here $\varepsilon_{2D}$ is obtained from a separate DFTB calculation, while other parameters are given directly by the structure (Fig.~\ref{fig:systems}). The rough agreement between this model and the calculated edge energies provides a familiar interpretation for the origin of edge energy as edge atom undercoordination (inset in Fig.~\ref{fig:edge_e}).

Next, we look at the spatial contributions. The edge energy contributions are the largest right at the edges, but they die out within a few lattice constants: edge energy penetrates the bulk around half a nanometer (Fig.~\ref{fig:edge_e}). As shown above, it is not surprising that the most important contributions to the edge energies come from the undercoordinated edge atoms. However, it is curious to see that although edge energies are positive at the system level, all edges also have at least one atom whose contribution is negative. These atoms have cohesion energy higher than 2D bulk atoms and are usually the nearest neighbors to the edge atoms. This higher cohesion presumably arises due to the edge atom, which can bind stronger to its neighbors due to the absence of competing bonds on the other side. Deeper in the bulk ($>0.5$~nm), the atoms become more bulk-like, with edge energy contributions converging rapidly toward zero. The studied materials and edges have no exceptions in this regard.

\subsection{Edge stress}
\label{sec:stress}
Unlike the edge energy, the system-level edge stress can be either positive or negative (Table~\ref{tab:data}). Ribbon tends to contract under positive (tensile) edge stress ($\tau>0$) and elongate under negative (compressive) edge stress ($\tau<0$). Au/sta, Au/str, hBN/z/N, and MoS$_2$/zz/S2 have tensile stress, while Gr/ac, Gr/zz, hBN/ac, hBN/zz/B, and MoS$_2$/zz/S have compressive edge stresses. These results generally agree with previous calculations, many of which have reported compressive edge stress for graphene edges \cite{Huang2009, Gan2010, Reddy2009}. Goldene has a tensile edge stress, in agreement with density-functional simulations \cite{abidi2025}. For hBN/ac and hBN/zz Huang \emph{et al.} report tensile edge stresses \cite{Huang2012}, while for hBN/zz/B and hBN/zz/N Yang \emph{et al.} report compressive edge stresses \cite{Yang2021}. Our numbers fall somewhere in between, hBN/zz/N having tensile, other hBN edges compressive stress. Qi \emph{et al.} reported tensile stress for MoS$_2$, but their result was a sum of both MoS$_2$/zz/S and MoS$_2$/zz/Mo edges, the latter of which was excluded from our set of edges.

\begin{figure}[t!]
    \centering
    \includegraphics[width=\linewidth]{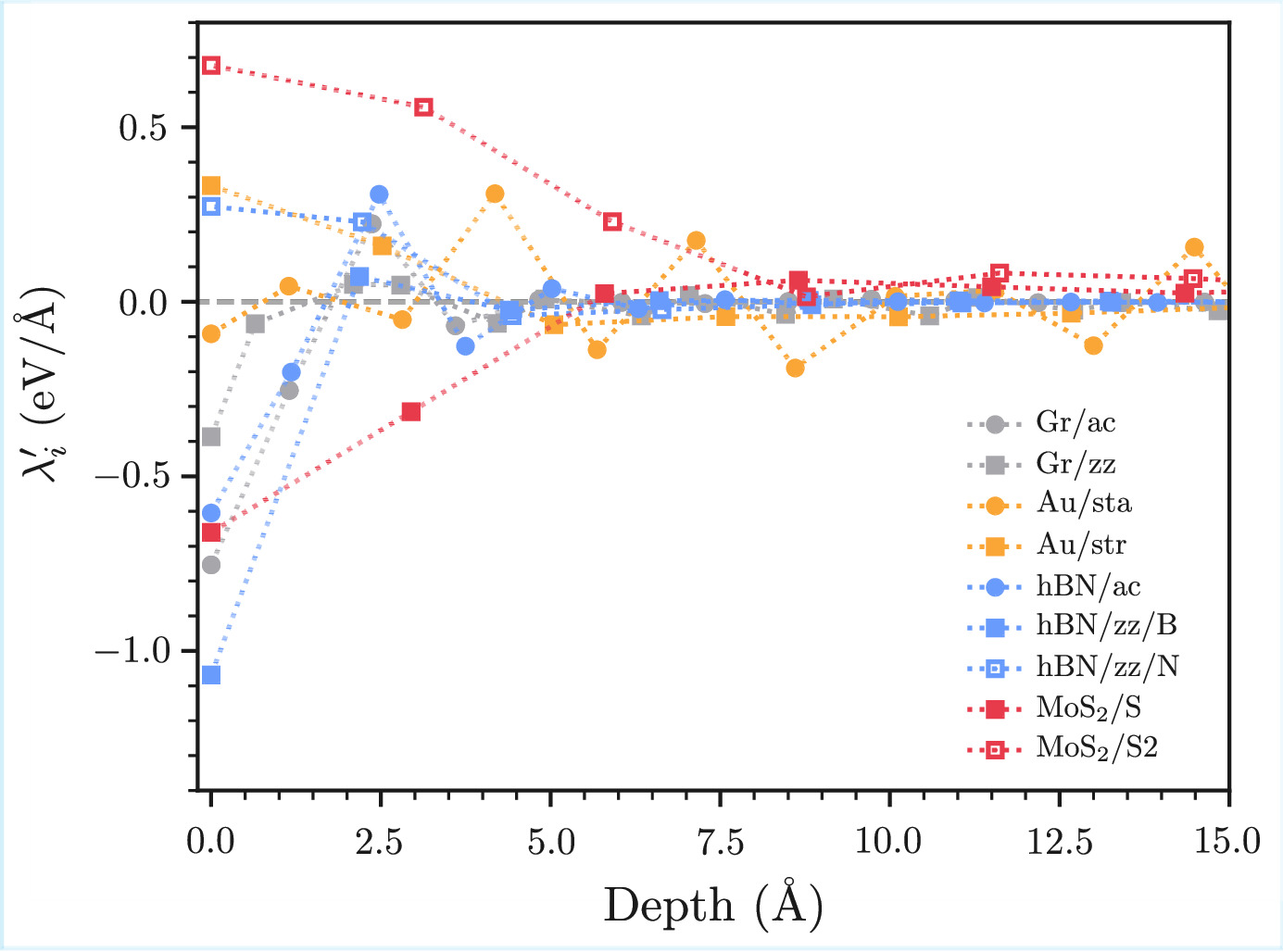}
    \caption{Spatial contributions to edge stresses as a function of bulk depth.}
    \label{fig:tau}
\end{figure} 

Beyond the system-level properties, for most edges, the spatial contributions for the edge stress penetrate the bulk less than a nanometer (Fig.~\ref{fig:tau}). A striking exception is goldene with a staggered edge. Its edge atoms contribute little, but contributions deeper in the bulk strengthen and fluctuate---practically even across the widest $5$-nm ribbon. Goldene with a straight edge does not share this behavior, as it, too, has dominant edge contributions and penetrates the bulk less than half a nanometer. 

The fluctuations of stress contributions in Au/sta may be due to quantum effects causing slight charge inhomogeneities, which in turn cause long-range effects through Coulomb interactions. These charge inhomogeneities are observable in the Mulliken charges, which fluctuate slightly for Au/sta edge but not for the other edges.
 
The stress contributions would look different if we used atomic resolution for hBN and MoS$_2$. In the bulk, B atoms have positive and N atoms equally negative edge stress contributions ($\tau_B+\tau_N=0)$. Similarly, the bulk Mo atoms have positive and bulk S atoms half of that negative edge stress contributions ($\tau_{Mo}+2\tau_S=0)$. This way, there are opposite energetic trends within the structural unit but not between structural units, so structural units do not contribute to the stress in the bulk. 

\subsection{Edge elastic modulus}
\label{sec:Ye}

The edge elastic moduli at the system level are either positive or negative and \revise{depend on the material and the edge} (Table~\ref{tab:data}). Here, the edges of Gr/zz, Au/str, hBN/zz/B, hBN/zz/N, and MoS$_2$/zz/S have positive and the other edges negative elastic moduli. This means that the edges make the ribbon stiffer for the former edges and looser for the latter edges; the effect is still small, with $|w_e|\lesssim 1$~\AA. Here $w_e=Y_e/Y$ measures the effective change in ribbon width (at one edge) required to produce the same effect as the edge elastic modulus.  

\label{ss:e_e}
\begin{figure}[t!]
    \centering
    \includegraphics[width=\linewidth]{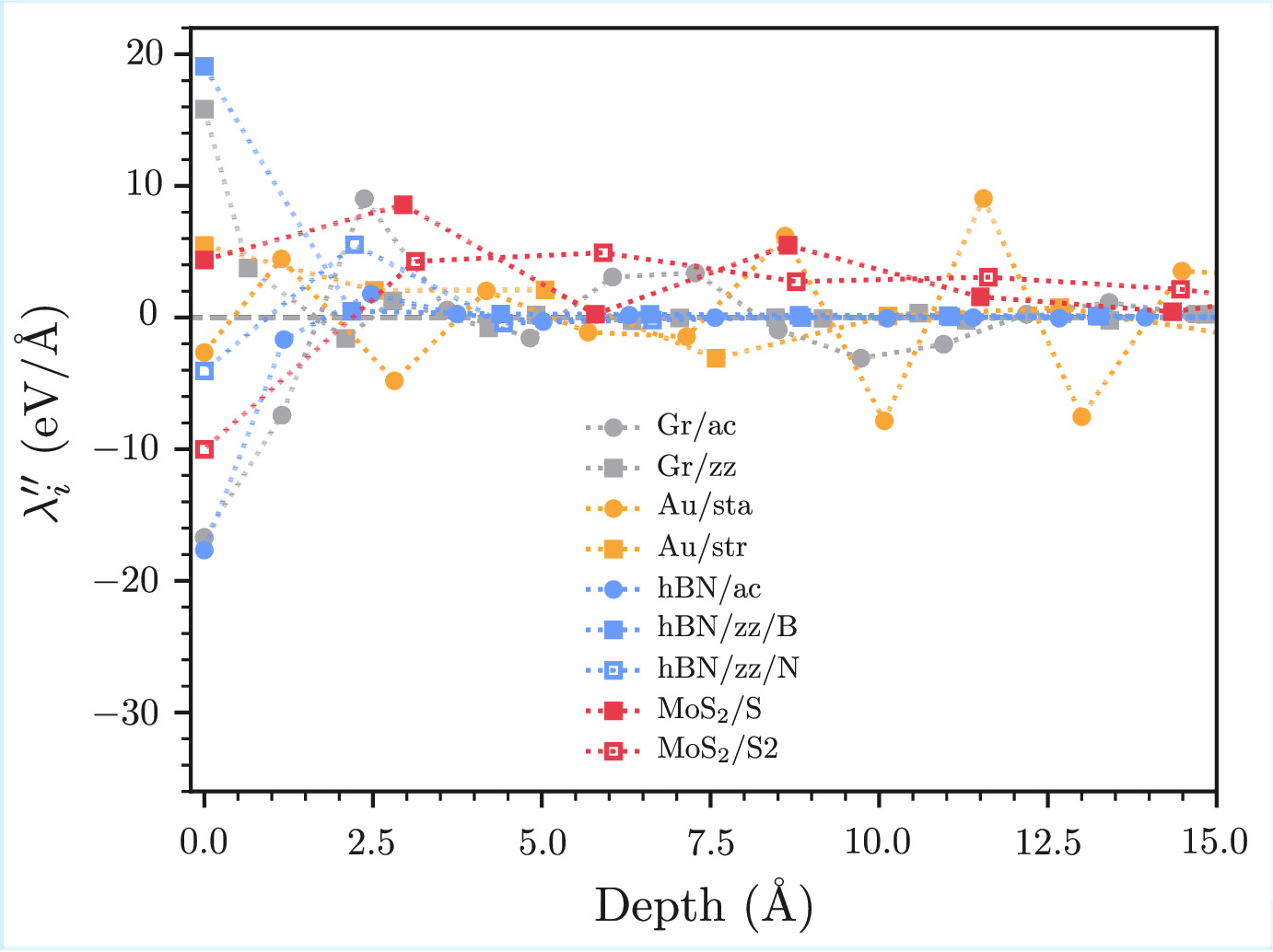}
    \caption{Spatial contributions to edge elastic moduli as a function of bulk depth.}
    \label{fig:edge_y}
\end{figure} 

Comparing elastic moduli to values in literature is more difficult than for energies and stresses. The difficulty arises because the edge elastic modulus depends on the choice of the width $w$ through Eq.~(\ref{eq:elastic}). Some studies have reported the edge elastic modulus without describing the choices made \cite{Gan2010, Huang2009}. Using the conventional choice for the width $w$, the values for Young's modulus follow trends reported in the literature. It is the highest for graphene and hBN, the lowest for goldene and MoS$_2$ (Table~\ref{tab:data}) \cite{Changgu2008, Falin2017, Mortazavi2024, Bertolazzi2011}. The edge elastic moduli are positive for Gr/ac, Gr/zz, and hBN/zz, with the highest modulus for Gr/zz and the lowest for Gr/ac, in agreement with the literature \cite{Huang2009, Reddy2009, Yang2021}. Literature values for $Y_e$ vary greatly also due to their numerical sensitivity as the second energy derivative. With scarce data, the second derivative and the fitting of Eq.~(\ref{eq:Eperl}) soon becomes numerically error-prone \cite{Huang2009}. Scarcity is a problem typically in expensive DFT calculations; we paid special attention to produce DFTB data abundantly and to ensure a numerically robust fit.

Spatial analysis shows that edge elastic modulus decays gradually but penetrates the bulk far deeper than energy and stress contributions (Fig.~\ref{fig:edge_y}). Also, the general scatter of the contributions is greater than that of previous edge properties. (The scatter is not due to numerical inaccuracy as noted above, it is a real edge property.) The penetration is generally above one nanometer and depends on the material and edge. The penetration below $0.5$~nm for Gr/zz and all edges of hBN, but above $\sim 1$~nm for other materials. It is peculiar how atoms nanometer-deep behave bulk-like in terms of energy and stress, but not in terms of elastic properties.

Once more, an exception to the steady decay of Young's modulus is goldene with a staggered edge. The spatial contributions fluctuate wildly across the ribbon, likely for the same reason as with stress (Sec.~\ref{sec:stress}). Thus, at least for Au/sta, the edge elastic modulus is really not an edge property but a property of the entire ribbon. 

\subsection{Electronic structure}

\begin{figure*}
    \centering
    \includegraphics[width=\linewidth]{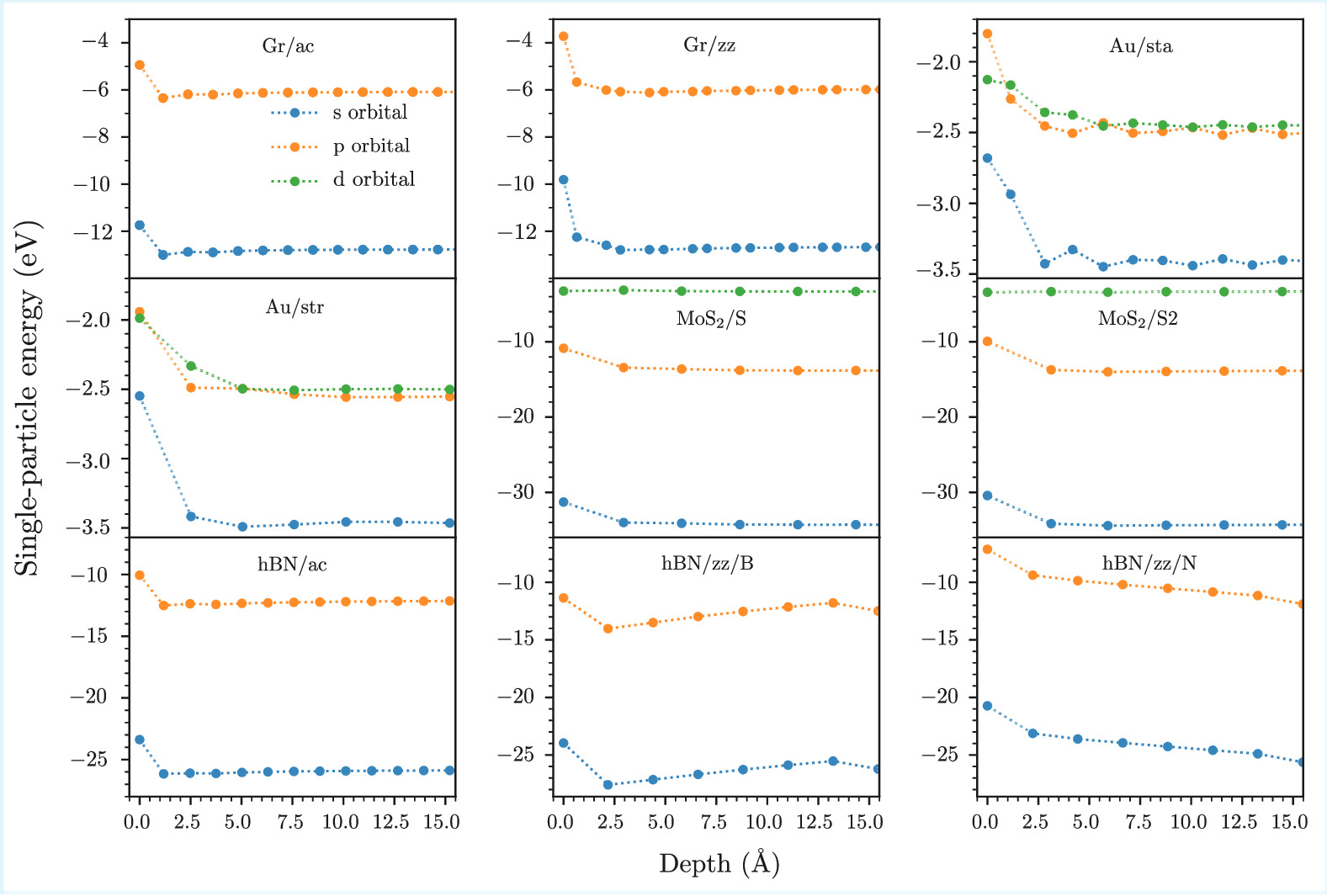}
    \caption{The mean energies of the orbital-projected local densities of states as a function of bulk depth. The local densities are calculated for structural units.}
    \label{fig:ldos}
\end{figure*} 

Finally, for a wider perspective, we extend the discussion to electronic structure properties. We investigated the effect of edges on the electronic structure through projected and averaged local density of states (LDOS). We define it for atom $i$ as
\begin{equation}
\epsilon_i^l = \frac{\int_{-\infty}^\infty f(\epsilon)\textrm{LDOS}_i^l(\epsilon)\epsilon d\epsilon}{\int_{-\infty}^\infty f(\epsilon)\textrm{LDOS}_i^l(\epsilon)d\epsilon}, 
\end{equation}
where $\epsilon$ is the single-particle energy, $f(\epsilon)$ is the Fermi function, LDOS$_i^l(\epsilon)$ is the density of states at energy $\epsilon$ projected to atom $i$ and angular momentum $l=0,1,2$. The Fermi broadening was $0.05$~eV. All materials had $s$- and $p$-projections, Au and Mo also $d$-projections. 

Overall, electronic structures converge to bulk values more rapidly than mechanical edge properties (Fig.~\ref{fig:ldos}). Penetration depths remain well below half a nanometer when quantified through this averaged property. These depths also provide the length scale within which the edge is expected to influence chemical properties such as adsorption energies on top of the 2D material.

The rapid convergence is defied by two notable exceptions: Au/sta and hBN/zz. For Au/sta $\epsilon_i^l$'s fluctuate somewhat across the ribbon, just like edge stress and edge elastic modulus contributions. These fluctuations further confirm that the behavior of Au/sta mechanical edge properties discussed earlier was not just due to noisy numerical differentiation. For hBN/zz/B $\epsilon_i^l$'s increase and for hBN/zz/N decrease when moving away from the edge. This opposite behavior arises from the asymmetric structure of the hBN ribbon. For all hBN/zz ribbons, one edge is hBN/zz/N (negatively charged edge atom) and the other end is hBN/zz/B (positively charged edge atom). Due to the extended range of Coulomb interaction, infeasibly wide hBN/zz ribbons would be required to get the electronic properties in the middle of the ribbon to plateau. Still, the mechanical properties did not show this asymmetry of hBN/zz edges. This feature is interesting because edge stresses and elastic moduli for other materials still penetrate the bulk less.

\section{Summary and conclusions}
In this article, we studied the mechanical edge properties, edge energy, edge stress, and edge elastic modulus beyond the system level by resolving \revise{their bulk penetration lengths}. The investigation of mechanical properties was augmented by spatial analysis of electronic structure properties. We studied nine edges of four 2D materials: graphene, goldene, hexagonal boron nitride, and molybdenum disulfide. Edge energy and edge stress contributions penetrated the bulk less than half a nanometer, but edge elastic moduli more than twice as deep. There was some variation between different edges. One exception was goldene with a staggered edge, where edge stress and edge elastic modulus contributions, as well as electronic structure properties, fluctuated across the entire ribbon. The exceptional behavior presumably arose from charge inhomogeneities due to quantum effects. Why these effects are visible only in Au/sta and not in Au/str or other edges remains an open question addressed in subsequent studies.

\revise{Our study has certain limitations. In addition to quantitative uncertainties related to the DFTB method itself, the nine different edges are not generic and cannot represent edges with passivating elements \cite{koskinen2008self}, edge roughness \cite{basu2008effect}, or curved geometries \cite{girit2009graphene}. For supported 2D materials, the edge properties may also be affected by substrate-induced charge transfer, not considered here \cite{yin2009bright}.}

As a takeaway message, \revise{in alignment with previous computational and experimental results on other edge-related properties \cite{tao2011spatially}, our study cautions against taking mechanical, system-level edge properties literally. The system-level mechanical properties of edge stresses and edge elastic moduli cannot be assumed to localize right at the edge; their edge penetration length must be considered.} Relaxing this assumption implies that care must be taken when modeling mechanical edge properties in continuum models. \revise{For example, typically the mechanical behavior of nanoribbons or edges of 2D materials would be modeled by using ribbon width $w$, Young's modulus $Y$, Poisson ratio $\sigma$, and bending modulus $B$, while assigning edge stress $\tau$ and edge elastic modulus $Y_e$ to be located \emph{right} at the edge \cite{shenoy2010spontaneous,EdgeStressInducedSpontaneous20212}. Because the strains in edge wrinkling and ribbon warping are inhomogeneous, assigning properties at the edge might induce these deformations, while assigning properties to be offset from the edge might not. Using such unfounded, erroneous assumptions may result in qualitatively wrong mechanical behavior.}

\subsection*{Appendix: Local energy expression in DFTB}
\revise{
Following Ref.~\onlinecite{hotbit}, the total DFTB energy of Eq.~(\ref{eq:edftb}) can be expressed as
\begin{equation}
E_{DFTB}=\sum_{i=1}^N e_i = \sum_{i=1}^N \left(A_i+\frac{1}{2}\sum_{j\neq i} B_{ij} \right),
\end{equation}
where the local contributions $e_i$ can be expressed as the sum of on-site and two-site interaction terms. The term 
\begin{align}
A_i=\frac{1}{2}\gamma_{ii} \Delta q_i^2 + E^\textrm{prom}_i
\end{align}
describes the on-site energy contributions, where $\gamma_{ii}=U_i$ is the on-site Hubbard parameter related to Coulomb interaction, $\Delta q_i$ is the excess Mulliken charge, and $E_i^\textrm{prom}$ is the promotion energy of atom $i$. The term 
\begin{align}
\begin{split}
B_{ij} =& V^\textrm{rep}_{ij} + \gamma_{ij}\Delta q_i \Delta q_j+ \sum_{\mu \in i, \nu \in j} E^{BS}_{\mu \nu}
\end{split}
\end{align}
represents the interaction energy between atoms $i$ and $j$, where $V_{ij}^\textrm{rep}$ is the repulsion and $\gamma_{ij}\Delta q_i \Delta q_j$ the Coulomb energy between atoms $i$ and $j$, and $E^{BS}_{\mu \nu}$ is the band structure energy related to orbitals $\mu$ on atom $i$ and $\nu$ on atom $j$. The expression for $e_i$, which arises naturally from the exact DFTB energy expression, uniquely describes atom $i$'s contribution to energy---in electrostatic, repulsive, promotional, and bonding sense.}

\begin{acknowledgments}
We acknowledge the Jane and Aatos Erkko Foundation for funding (project EcoMet) and the Finnish Grid and Cloud Infrastructure (FGCI) and CSC—IT Center for Science for computational resources.
\end{acknowledgments}


\begin{thebibliography}{55}%
	\makeatletter
	\providecommand \@ifxundefined [1]{%
		\@ifx{#1\undefined}
	}%
	\providecommand \@ifnum [1]{%
		\ifnum #1\expandafter \@firstoftwo
		\else \expandafter \@secondoftwo
		\fi
	}%
	\providecommand \@ifx [1]{%
		\ifx #1\expandafter \@firstoftwo
		\else \expandafter \@secondoftwo
		\fi
	}%
	\providecommand \natexlab [1]{#1}%
	\providecommand \enquote  [1]{``#1''}%
	\providecommand \bibnamefont  [1]{#1}%
	\providecommand \bibfnamefont [1]{#1}%
	\providecommand \citenamefont [1]{#1}%
	\providecommand \href@noop [0]{\@secondoftwo}%
	\providecommand \href [0]{\begingroup \@sanitize@url \@href}%
	\providecommand \@href[1]{\@@startlink{#1}\@@href}%
	\providecommand \@@href[1]{\endgroup#1\@@endlink}%
	\providecommand \@sanitize@url [0]{\catcode `\\12\catcode `\$12\catcode
		`\&12\catcode `\#12\catcode `\^12\catcode `\_12\catcode `\%12\relax}%
	\providecommand \@@startlink[1]{}%
	\providecommand \@@endlink[0]{}%
	\providecommand \url  [0]{\begingroup\@sanitize@url \@url }%
	\providecommand \@url [1]{\endgroup\@href {#1}{\urlprefix }}%
	\providecommand \urlprefix  [0]{URL }%
	\providecommand \Eprint [0]{\href }%
	\providecommand \doibase [0]{https://doi.org/}%
	\providecommand \selectlanguage [0]{\@gobble}%
	\providecommand \bibinfo  [0]{\@secondoftwo}%
	\providecommand \bibfield  [0]{\@secondoftwo}%
	\providecommand \translation [1]{[#1]}%
	\providecommand \BibitemOpen [0]{}%
	\providecommand \bibitemStop [0]{}%
	\providecommand \bibitemNoStop [0]{.\EOS\space}%
	\providecommand \EOS [0]{\spacefactor3000\relax}%
	\providecommand \BibitemShut  [1]{\csname bibitem#1\endcsname}%
	\let\auto@bib@innerbib\@empty
	\bibitem [{\citenamefont {Jia}\ \emph {et~al.}(2011)\citenamefont {Jia},
		\citenamefont {Campos-Delgado}, \citenamefont {Terrones}, \citenamefont
		{Meunier},\ and\ \citenamefont {Dresselhaus}}]{graphene_edge_review}%
	\BibitemOpen
	\bibfield  {author} {\bibinfo {author} {\bibfnamefont {X.}~\bibnamefont
			{Jia}}, \bibinfo {author} {\bibfnamefont {J.}~\bibnamefont {Campos-Delgado}},
		\bibinfo {author} {\bibfnamefont {M.}~\bibnamefont {Terrones}}, \bibinfo
		{author} {\bibfnamefont {V.}~\bibnamefont {Meunier}},\ and\ \bibinfo {author}
		{\bibfnamefont {M.~S.}\ \bibnamefont {Dresselhaus}},\ }\bibfield  {title}
	{\bibinfo {title} {Graphene edges: a review of their fabrication and
			characterization},\ }\href@noop {} {\bibfield  {journal} {\bibinfo  {journal}
			{Nanoscale}\ }\textbf {\bibinfo {volume} {3}},\ \bibinfo {pages} {86}
		(\bibinfo {year} {2011})}\BibitemShut {NoStop}%
	\bibitem [{\citenamefont {Li}\ \emph {et~al.}(2023)\citenamefont {Li},
		\citenamefont {Yin}, \citenamefont {Gao}, \citenamefont {Guo}, \citenamefont
		{Zhao}, \citenamefont {Jia},\ and\ \citenamefont {Guo}}]{edge_review_newer}%
	\BibitemOpen
	\bibfield  {author} {\bibinfo {author} {\bibfnamefont {M.}~\bibnamefont
			{Li}}, \bibinfo {author} {\bibfnamefont {B.}~\bibnamefont {Yin}}, \bibinfo
		{author} {\bibfnamefont {C.}~\bibnamefont {Gao}}, \bibinfo {author}
		{\bibfnamefont {J.}~\bibnamefont {Guo}}, \bibinfo {author} {\bibfnamefont
			{C.}~\bibnamefont {Zhao}}, \bibinfo {author} {\bibfnamefont {C.}~\bibnamefont
			{Jia}},\ and\ \bibinfo {author} {\bibfnamefont {X.}~\bibnamefont {Guo}},\
	}\bibfield  {title} {\bibinfo {title} {Graphene: Preparation, tailoring, and
			modification},\ }\href {https://doi.org/https://doi.org/10.1002/EXP.20210233}
	{\bibfield  {journal} {\bibinfo  {journal} {Exploration}\ }\textbf {\bibinfo
			{volume} {3}},\ \bibinfo {pages} {20210233} (\bibinfo {year}
		{2023})}\BibitemShut {NoStop}%
	\bibitem [{\citenamefont {Berashevich}\ and\ \citenamefont
		{Chakraborty}(2009)}]{adsorption1}%
	\BibitemOpen
	\bibfield  {author} {\bibinfo {author} {\bibfnamefont {J.}~\bibnamefont
			{Berashevich}}\ and\ \bibinfo {author} {\bibfnamefont {T.}~\bibnamefont
			{Chakraborty}},\ }\bibfield  {title} {\bibinfo {title} {Tunable band gap and
			magnetic ordering by adsorption of molecules on graphene},\ }\href
	{https://doi.org/10.1103/PhysRevB.80.033404} {\bibfield  {journal} {\bibinfo
			{journal} {Phys. Rev. B}\ }\textbf {\bibinfo {volume} {80}},\ \bibinfo
		{pages} {033404} (\bibinfo {year} {2009})}\BibitemShut {NoStop}%
	\bibitem [{\citenamefont {Uthaisar}\ \emph {et~al.}(2009)\citenamefont
		{Uthaisar}, \citenamefont {Barone},\ and\ \citenamefont
		{Peralta}}]{adsorption2}%
	\BibitemOpen
	\bibfield  {author} {\bibinfo {author} {\bibfnamefont {C.}~\bibnamefont
			{Uthaisar}}, \bibinfo {author} {\bibfnamefont {V.}~\bibnamefont {Barone}},\
		and\ \bibinfo {author} {\bibfnamefont {J.~E.}\ \bibnamefont {Peralta}},\
	}\bibfield  {title} {\bibinfo {title} {Lithium adsorption on zigzag graphene
			nanoribbons},\ }\href {https://doi.org/10.1063/1.3265431} {\bibfield
		{journal} {\bibinfo  {journal} {J. Appl. Phys.}\ }\textbf {\bibinfo {volume}
			{106}},\ \bibinfo {pages} {113715} (\bibinfo {year} {2009})}\BibitemShut
	{NoStop}%
	\bibitem [{\citenamefont {Kumar}\ \emph {et~al.}(2023)\citenamefont {Kumar},
		\citenamefont {Pratap}, \citenamefont {Kumar}, \citenamefont {Mishra},
		\citenamefont {Gwag},\ and\ \citenamefont {Chakraborty}}]{sensors}%
	\BibitemOpen
	\bibfield  {author} {\bibinfo {author} {\bibfnamefont {S.}~\bibnamefont
			{Kumar}}, \bibinfo {author} {\bibfnamefont {S.}~\bibnamefont {Pratap}},
		\bibinfo {author} {\bibfnamefont {V.}~\bibnamefont {Kumar}}, \bibinfo
		{author} {\bibfnamefont {R.~K.}\ \bibnamefont {Mishra}}, \bibinfo {author}
		{\bibfnamefont {J.~S.}\ \bibnamefont {Gwag}},\ and\ \bibinfo {author}
		{\bibfnamefont {B.}~\bibnamefont {Chakraborty}},\ }\bibfield  {title}
	{\bibinfo {title} {Electronic, transport, magnetic, and optical properties of
			graphene nanoribbons and their optical sensing applications: A comprehensive
			review},\ }\href {https://doi.org/https://doi.org/10.1002/bio.4334}
	{\bibfield  {journal} {\bibinfo  {journal} {Luminescence}\ }\textbf {\bibinfo
			{volume} {38}},\ \bibinfo {pages} {909953} (\bibinfo {year}
		{2023})}\BibitemShut {NoStop}%
	\bibitem [{\citenamefont {Solis-Fernandez}\ \emph {et~al.}(2017)\citenamefont
		{Solis-Fernandez}, \citenamefont {Bissett},\ and\ \citenamefont
		{Ago}}]{Heterostructure_application}%
	\BibitemOpen
	\bibfield  {author} {\bibinfo {author} {\bibfnamefont {P.}~\bibnamefont
			{Solís-Fernández}}, \bibinfo {author} {\bibfnamefont {M.}~\bibnamefont
			{Bissett}},\ and\ \bibinfo {author} {\bibfnamefont {H.}~\bibnamefont {Ago}},\
	}\bibfield  {title} {\bibinfo {title} {Synthesis{,} structure and
			applications of graphene-based {2D} heterostructures},\ }\href
	{https://doi.org/10.1039/C7CS00160F} {\bibfield  {journal} {\bibinfo
			{journal} {Chem. Soc. Rev.}\ }\textbf {\bibinfo {volume} {46}},\ \bibinfo
		{pages} {45724613} (\bibinfo {year} {2017})}\BibitemShut {NoStop}%
	\bibitem [{\citenamefont {Ni}\ and\ \citenamefont {Wang}(2015)}]{catalysis}%
	\BibitemOpen
	\bibfield  {author} {\bibinfo {author} {\bibfnamefont {B.}~\bibnamefont
			{Ni}}\ and\ \bibinfo {author} {\bibfnamefont {X.}~\bibnamefont {Wang}},\
	}\bibfield  {title} {\bibinfo {title} {Face the edges: Catalytic active sites
			of nanomaterials},\ }\href
	{https://doi.org/https://doi.org/10.1002/advs.201500085} {\bibfield
		{journal} {\bibinfo  {journal} {Adv. Sci.}\ }\textbf {\bibinfo {volume}
			{2}},\ \bibinfo {pages} {1500085} (\bibinfo {year} {2015})}\BibitemShut
	{NoStop}%
	\bibitem [{\citenamefont {Xu}\ \emph {et~al.}(2021)\citenamefont {Xu},
		\citenamefont {Zhu}, \citenamefont {Ma}, \citenamefont {Ma}, \citenamefont
		{Bai}, \citenamefont {Chen},\ and\ \citenamefont {Mu}}]{MoS2_synthesis}%
	\BibitemOpen
	\bibfield  {author} {\bibinfo {author} {\bibfnamefont {H.}~\bibnamefont
			{Xu}}, \bibinfo {author} {\bibfnamefont {J.}~\bibnamefont {Zhu}}, \bibinfo
		{author} {\bibfnamefont {Q.}~\bibnamefont {Ma}}, \bibinfo {author}
		{\bibfnamefont {J.}~\bibnamefont {Ma}}, \bibinfo {author} {\bibfnamefont
			{H.}~\bibnamefont {Bai}}, \bibinfo {author} {\bibfnamefont {L.}~\bibnamefont
			{Chen}},\ and\ \bibinfo {author} {\bibfnamefont {S.}~\bibnamefont {Mu}},\
	}\bibfield  {title} {\bibinfo {title} {Two-dimensional {MoS}$_2$: Structural
			properties, synthesis methods, and regulation strategies toward oxygen
			reduction},\ }\href {https://www.mdpi.com/2072-666X/12/3/240} {\bibfield
		{journal} {\bibinfo  {journal} {Micromachines}\ }\textbf {\bibinfo {volume}
			{12}} (\bibinfo {year} {2021})}\BibitemShut {NoStop}%
	\bibitem [{\citenamefont {Ruffieux}\ \emph {et~al.}(2016)\citenamefont
		{Ruffieux}, \citenamefont {Wang}, \citenamefont {Yang}, \citenamefont
		{Sanchez-Sanchez}, \citenamefont {Liu}, \citenamefont {Dienel},
		\citenamefont {Talirz}, \citenamefont {Shinde}, \citenamefont {Pignedoli},
		\citenamefont {Passerone}, \citenamefont {Dumslaff}, \citenamefont {Feng},
		\citenamefont {Müllen},\ and\ \citenamefont {Fasel}}]{synthesis}%
	\BibitemOpen
	\bibfield  {author} {\bibinfo {author} {\bibfnamefont {P.}~\bibnamefont
			{Ruffieux}}, \bibinfo {author} {\bibfnamefont {S.}~\bibnamefont {Wang}},
		\bibinfo {author} {\bibfnamefont {B.}~\bibnamefont {Yang}}, \bibinfo {author}
		{\bibfnamefont {C.}~\bibnamefont {Sánchez-Sánchez}}, \bibinfo {author}
		{\bibfnamefont {J.}~\bibnamefont {Liu}}, \bibinfo {author} {\bibfnamefont
			{T.}~\bibnamefont {Dienel}}, \bibinfo {author} {\bibfnamefont
			{L.}~\bibnamefont {Talirz}}, \bibinfo {author} {\bibfnamefont
			{P.}~\bibnamefont {Shinde}}, \bibinfo {author} {\bibfnamefont {C.~A.}\
			\bibnamefont {Pignedoli}}, \bibinfo {author} {\bibfnamefont {D.}~\bibnamefont
			{Passerone}}, \bibinfo {author} {\bibfnamefont {T.}~\bibnamefont {Dumslaff}},
		\bibinfo {author} {\bibfnamefont {X.}~\bibnamefont {Feng}}, \bibinfo {author}
		{\bibfnamefont {K.}~\bibnamefont {Müllen}},\ and\ \bibinfo {author}
		{\bibfnamefont {R.}~\bibnamefont {Fasel}},\ }\bibfield  {title} {\bibinfo
		{title} {On-surface synthesis of graphene nanoribbons with zigzag edge
			topology},\ }\href {https://doi.org/10.1038/nature17151} {\bibfield
		{journal} {\bibinfo  {journal} {Nature}\ }\textbf {\bibinfo {volume} {531}},\
		\bibinfo {pages} {489} (\bibinfo {year} {2016})}\BibitemShut {NoStop}%
	\bibitem [{\citenamefont {Dean}\ \emph {et~al.}(2012)\citenamefont {Dean},
		\citenamefont {Young}, \citenamefont {Wang}, \citenamefont {Meric},
		\citenamefont {Lee}, \citenamefont {Watanabe}, \citenamefont {Taniguchi},
		\citenamefont {Shepard}, \citenamefont {Kim},\ and\ \citenamefont
		{Hone}}]{heterostructure}%
	\BibitemOpen
	\bibfield  {author} {\bibinfo {author} {\bibfnamefont {C.}~\bibnamefont
			{Dean}}, \bibinfo {author} {\bibfnamefont {A.}~\bibnamefont {Young}},
		\bibinfo {author} {\bibfnamefont {L.}~\bibnamefont {Wang}}, \bibinfo {author}
		{\bibfnamefont {I.}~\bibnamefont {Meric}}, \bibinfo {author} {\bibfnamefont
			{G.-H.}\ \bibnamefont {Lee}}, \bibinfo {author} {\bibfnamefont
			{K.}~\bibnamefont {Watanabe}}, \bibinfo {author} {\bibfnamefont
			{T.}~\bibnamefont {Taniguchi}}, \bibinfo {author} {\bibfnamefont
			{K.}~\bibnamefont {Shepard}}, \bibinfo {author} {\bibfnamefont
			{P.}~\bibnamefont {Kim}},\ and\ \bibinfo {author} {\bibfnamefont
			{J.}~\bibnamefont {Hone}},\ }\bibfield  {title} {\bibinfo {title} {Graphene
			based heterostructures},\ }\href
	{https://doi.org/https://doi.org/10.1016/j.ssc.2012.04.021} {\bibfield
		{journal} {\bibinfo  {journal} {Solid State Commun.}\ }\textbf {\bibinfo
			{volume} {152}},\ \bibinfo {pages} {1275} (\bibinfo {year}
		{2012})}\BibitemShut {NoStop}%
	\bibitem [{\citenamefont {Tao}\ \emph {et~al.}(2011)\citenamefont {Tao},
		\citenamefont {Jiao}, \citenamefont {Yazyev}, \citenamefont {Chen},
		\citenamefont {Feng}, \citenamefont {Zhang}, \citenamefont {Capaz},
		\citenamefont {Tour}, \citenamefont {Zettl}, \citenamefont {Louie} \emph
		{et~al.}}]{tao2011spatially}%
	\BibitemOpen
	\bibfield  {author} {\bibinfo {author} {\bibfnamefont {C.}~\bibnamefont
			{Tao}}, \bibinfo {author} {\bibfnamefont {L.}~\bibnamefont {Jiao}}, \bibinfo
		{author} {\bibfnamefont {O.~V.}\ \bibnamefont {Yazyev}}, \bibinfo {author}
		{\bibfnamefont {Y.-C.}\ \bibnamefont {Chen}}, \bibinfo {author}
		{\bibfnamefont {J.}~\bibnamefont {Feng}}, \bibinfo {author} {\bibfnamefont
			{X.}~\bibnamefont {Zhang}}, \bibinfo {author} {\bibfnamefont {R.~B.}\
			\bibnamefont {Capaz}}, \bibinfo {author} {\bibfnamefont {J.~M.}\ \bibnamefont
			{Tour}}, \bibinfo {author} {\bibfnamefont {A.}~\bibnamefont {Zettl}},
		\bibinfo {author} {\bibfnamefont {S.~G.}\ \bibnamefont {Louie}}, \emph
		{et~al.},\ }\bibfield  {title} {\bibinfo {title} {Spatially resolving edge
			states of chiral graphene nanoribbons},\ }\href@noop {} {\bibfield  {journal}
		{\bibinfo  {journal} {Nat. Phys.}\ }\textbf {\bibinfo {volume} {7}},\
		\bibinfo {pages} {616} (\bibinfo {year} {2011})}\BibitemShut {NoStop}%
	\bibitem [{\citenamefont {Fujita}\ \emph {et~al.}(1996)\citenamefont {Fujita},
		\citenamefont {Wakabayashi}, \citenamefont {Nakada},\ and\ \citenamefont
		{Kusakabe}}]{fujita1996peculiar}%
	\BibitemOpen
	\bibfield  {author} {\bibinfo {author} {\bibfnamefont {M.}~\bibnamefont
			{Fujita}}, \bibinfo {author} {\bibfnamefont {K.}~\bibnamefont {Wakabayashi}},
		\bibinfo {author} {\bibfnamefont {K.}~\bibnamefont {Nakada}},\ and\ \bibinfo
		{author} {\bibfnamefont {K.}~\bibnamefont {Kusakabe}},\ }\bibfield  {title}
	{\bibinfo {title} {Peculiar localized state at zigzag graphite edge},\
	}\href@noop {} {\bibfield  {journal} {\bibinfo  {journal} {J. Phys. Soc.
				Jpn.}\ }\textbf {\bibinfo {volume} {65}},\ \bibinfo {pages} {1920} (\bibinfo
		{year} {1996})}\BibitemShut {NoStop}%
	\bibitem [{\citenamefont {Kobayashi}\ \emph {et~al.}(2006)\citenamefont
		{Kobayashi}, \citenamefont {Fukui}, \citenamefont {Enoki},\ and\
		\citenamefont {Kusakabe}}]{kobayashi2006edge}%
	\BibitemOpen
	\bibfield  {author} {\bibinfo {author} {\bibfnamefont {Y.}~\bibnamefont
			{Kobayashi}}, \bibinfo {author} {\bibfnamefont {K.-i.}\ \bibnamefont
			{Fukui}}, \bibinfo {author} {\bibfnamefont {T.}~\bibnamefont {Enoki}},\ and\
		\bibinfo {author} {\bibfnamefont {K.}~\bibnamefont {Kusakabe}},\ }\bibfield
	{title} {\bibinfo {title} {Edge state on hydrogen-terminated graphite edges
			investigated by scanning tunneling microscopy},\ }\href@noop {} {\bibfield
		{journal} {\bibinfo  {journal} {Phys. Rev. B}\ }\textbf {\bibinfo {volume}
			{73}},\ \bibinfo {pages} {125415} (\bibinfo {year} {2006})}\BibitemShut
	{NoStop}%
	\bibitem [{\citenamefont {Wang}\ \emph {et~al.}(2016)\citenamefont {Wang},
		\citenamefont {Talirz}, \citenamefont {Pignedoli}, \citenamefont {Feng},
		\citenamefont {M{\"u}llen}, \citenamefont {Fasel},\ and\ \citenamefont
		{Ruffieux}}]{wang2016giant}%
	\BibitemOpen
	\bibfield  {author} {\bibinfo {author} {\bibfnamefont {S.}~\bibnamefont
			{Wang}}, \bibinfo {author} {\bibfnamefont {L.}~\bibnamefont {Talirz}},
		\bibinfo {author} {\bibfnamefont {C.~A.}\ \bibnamefont {Pignedoli}}, \bibinfo
		{author} {\bibfnamefont {X.}~\bibnamefont {Feng}}, \bibinfo {author}
		{\bibfnamefont {K.}~\bibnamefont {M{\"u}llen}}, \bibinfo {author}
		{\bibfnamefont {R.}~\bibnamefont {Fasel}},\ and\ \bibinfo {author}
		{\bibfnamefont {P.}~\bibnamefont {Ruffieux}},\ }\bibfield  {title} {\bibinfo
		{title} {Giant edge state splitting at atomically precise graphene zigzag
			edges},\ }\href@noop {} {\bibfield  {journal} {\bibinfo  {journal} {Nat.
				Commun.}\ }\textbf {\bibinfo {volume} {7}},\ \bibinfo {pages} {11507}
		(\bibinfo {year} {2016})}\BibitemShut {NoStop}%
	\bibitem [{\citenamefont {Bollinger}\ \emph {et~al.}(2001)\citenamefont
		{Bollinger}, \citenamefont {Lauritsen}, \citenamefont {Jacobsen},
		\citenamefont {N{\o}rskov}, \citenamefont {Helveg},\ and\ \citenamefont
		{Besenbacher}}]{bollinger2001one}%
	\BibitemOpen
	\bibfield  {author} {\bibinfo {author} {\bibfnamefont {M.}~\bibnamefont
			{Bollinger}}, \bibinfo {author} {\bibfnamefont {J.}~\bibnamefont
			{Lauritsen}}, \bibinfo {author} {\bibfnamefont {K.~W.}\ \bibnamefont
			{Jacobsen}}, \bibinfo {author} {\bibfnamefont {J.~K.}\ \bibnamefont
			{N{\o}rskov}}, \bibinfo {author} {\bibfnamefont {S.}~\bibnamefont {Helveg}},\
		and\ \bibinfo {author} {\bibfnamefont {F.}~\bibnamefont {Besenbacher}},\
	}\bibfield  {title} {\bibinfo {title} {One-dimensional metallic edge states
			in mos 2},\ }\href@noop {} {\bibfield  {journal} {\bibinfo  {journal} {Phys.
				Rev. Lett.}\ }\textbf {\bibinfo {volume} {87}},\ \bibinfo {pages} {196803}
		(\bibinfo {year} {2001})}\BibitemShut {NoStop}%
	\bibitem [{\citenamefont {Lauritsen}\ \emph {et~al.}(2004)\citenamefont
		{Lauritsen}, \citenamefont {Nyberg}, \citenamefont {N{\o}rskov},
		\citenamefont {Clausen}, \citenamefont {Tops{\o}e}, \citenamefont
		{L{\ae}gsgaard},\ and\ \citenamefont
		{Besenbacher}}]{lauritsen2004hydrodesulfurization}%
	\BibitemOpen
	\bibfield  {author} {\bibinfo {author} {\bibfnamefont {J.}~\bibnamefont
			{Lauritsen}}, \bibinfo {author} {\bibfnamefont {M.}~\bibnamefont {Nyberg}},
		\bibinfo {author} {\bibfnamefont {J.~K.}\ \bibnamefont {N{\o}rskov}},
		\bibinfo {author} {\bibfnamefont {B.}~\bibnamefont {Clausen}}, \bibinfo
		{author} {\bibfnamefont {H.}~\bibnamefont {Tops{\o}e}}, \bibinfo {author}
		{\bibfnamefont {E.}~\bibnamefont {L{\ae}gsgaard}},\ and\ \bibinfo {author}
		{\bibfnamefont {F.}~\bibnamefont {Besenbacher}},\ }\bibfield  {title}
	{\bibinfo {title} {Hydrodesulfurization reaction pathways on mos2
			nanoclusters revealed by scanning tunneling microscopy},\ }\href@noop {}
	{\bibfield  {journal} {\bibinfo  {journal} {Journal of Catalysis}\ }\textbf
		{\bibinfo {volume} {224}},\ \bibinfo {pages} {94} (\bibinfo {year}
		{2004})}\BibitemShut {NoStop}%
	\bibitem [{\citenamefont {Reddy}\ \emph {et~al.}(2009)\citenamefont {Reddy},
		\citenamefont {Ramasubramaniam}, \citenamefont {Shenoy},\ and\ \citenamefont
		{Zhang}}]{Reddy2009}%
	\BibitemOpen
	\bibfield  {author} {\bibinfo {author} {\bibfnamefont {C.~D.}\ \bibnamefont
			{Reddy}}, \bibinfo {author} {\bibfnamefont {A.}~\bibnamefont
			{Ramasubramaniam}}, \bibinfo {author} {\bibfnamefont {V.~B.}\ \bibnamefont
			{Shenoy}},\ and\ \bibinfo {author} {\bibfnamefont {Y.-W.}\ \bibnamefont
			{Zhang}},\ }\bibfield  {title} {\bibinfo {title} {Edge elastic properties of
			defect-free single-layer graphene sheets},\ }\href
	{https://doi.org/10.1063/1.3094878} {\bibfield  {journal} {\bibinfo
			{journal} {Appl. Phys. Lett.}\ }\textbf {\bibinfo {volume} {94}},\ \bibinfo
		{pages} {101904} (\bibinfo {year} {2009})}\BibitemShut {NoStop}%
	\bibitem [{\citenamefont {Gan}\ and\ \citenamefont
		{Srolovitz}(2010)}]{Gan2010}%
	\BibitemOpen
	\bibfield  {author} {\bibinfo {author} {\bibfnamefont {C.~K.}\ \bibnamefont
			{Gan}}\ and\ \bibinfo {author} {\bibfnamefont {D.~J.}\ \bibnamefont
			{Srolovitz}},\ }\bibfield  {title} {\bibinfo {title} {First-principles study
			of graphene edge properties and flake shapes},\ }\href
	{https://doi.org/10.1103/PhysRevB.81.125445} {\bibfield  {journal} {\bibinfo
			{journal} {Phys. Rev. B}\ }\textbf {\bibinfo {volume} {81}},\ \bibinfo
		{pages} {125445} (\bibinfo {year} {2010})}\BibitemShut {NoStop}%
	\bibitem [{\citenamefont {Huang}\ \emph {et~al.}(2009)\citenamefont {Huang},
		\citenamefont {Liu}, \citenamefont {Su}, \citenamefont {Wu}, \citenamefont
		{Duan}, \citenamefont {Gu},\ and\ \citenamefont {Liu}}]{Huang2009}%
	\BibitemOpen
	\bibfield  {author} {\bibinfo {author} {\bibfnamefont {B.}~\bibnamefont
			{Huang}}, \bibinfo {author} {\bibfnamefont {M.}~\bibnamefont {Liu}}, \bibinfo
		{author} {\bibfnamefont {N.}~\bibnamefont {Su}}, \bibinfo {author}
		{\bibfnamefont {J.}~\bibnamefont {Wu}}, \bibinfo {author} {\bibfnamefont
			{W.}~\bibnamefont {Duan}}, \bibinfo {author} {\bibfnamefont {B.-l.}\
			\bibnamefont {Gu}},\ and\ \bibinfo {author} {\bibfnamefont {F.}~\bibnamefont
			{Liu}},\ }\bibfield  {title} {\bibinfo {title} {Quantum manifestations of
			graphene edge stress and edge instability: A first-principles study},\ }\href
	{https://doi.org/10.1103/PhysRevLett.102.166404} {\bibfield  {journal}
		{\bibinfo  {journal} {Phys. Rev. Lett.}\ }\textbf {\bibinfo {volume} {102}},\
		\bibinfo {pages} {166404} (\bibinfo {year} {2009})}\BibitemShut {NoStop}%
	\bibitem [{\citenamefont {Abidi}\ \emph {et~al.}(2025)\citenamefont {Abidi},
		\citenamefont {Bagheri}, \citenamefont {Singh},\ and\ \citenamefont
		{Koskinen}}]{abidi2025}%
	\BibitemOpen
	\bibfield  {author} {\bibinfo {author} {\bibfnamefont {K.~R.}\ \bibnamefont
			{Abidi}}, \bibinfo {author} {\bibfnamefont {M.}~\bibnamefont {Bagheri}},
		\bibinfo {author} {\bibfnamefont {S.}~\bibnamefont {Singh}},\ and\ \bibinfo
		{author} {\bibfnamefont {P.}~\bibnamefont {Koskinen}},\ }\bibfield  {title}
	{\bibinfo {title} {Atomically thin metallenes at the edge},\ }\href@noop {}
	{\bibfield  {journal} {\bibinfo  {journal} {2D Mater.}\ }\textbf {\bibinfo
			{volume} {12}},\ \bibinfo {pages} {025016} (\bibinfo {year}
		{2025})}\BibitemShut {NoStop}%
	\bibitem [{\citenamefont {Huang}\ \emph {et~al.}(2012)\citenamefont {Huang},
		\citenamefont {Lee}, \citenamefont {Gu}, \citenamefont {Liu},\ and\
		\citenamefont {Duan}}]{Huang2012}%
	\BibitemOpen
	\bibfield  {author} {\bibinfo {author} {\bibfnamefont {B.}~\bibnamefont
			{Huang}}, \bibinfo {author} {\bibfnamefont {H.}~\bibnamefont {Lee}}, \bibinfo
		{author} {\bibfnamefont {B.-L.}\ \bibnamefont {Gu}}, \bibinfo {author}
		{\bibfnamefont {F.}~\bibnamefont {Liu}},\ and\ \bibinfo {author}
		{\bibfnamefont {W.}~\bibnamefont {Duan}},\ }\bibfield  {title} {\bibinfo
		{title} {Edge stability of boron nitride nanoribbons and its application in
			designing hybrid bnc structures},\ }\href
	{https://doi.org/10.1007/s12274-011-0185-y} {\bibfield  {journal} {\bibinfo
			{journal} {Nano Res.}\ }\textbf {\bibinfo {volume} {5}},\ \bibinfo {pages}
		{62} (\bibinfo {year} {2012})}\BibitemShut {NoStop}%
	\bibitem [{\citenamefont {Qi}\ \emph {et~al.}(2013)\citenamefont {Qi},
		\citenamefont {Cao},\ and\ \citenamefont {Park}}]{Qi2013}%
	\BibitemOpen
	\bibfield  {author} {\bibinfo {author} {\bibfnamefont {Z.}~\bibnamefont
			{Qi}}, \bibinfo {author} {\bibfnamefont {P.}~\bibnamefont {Cao}},\ and\
		\bibinfo {author} {\bibfnamefont {H.~S.}\ \bibnamefont {Park}},\ }\bibfield
	{title} {\bibinfo {title} {Density functional theory calculation of edge
			stresses in monolayer {MoS}$_2$},\ }\href {https://doi.org/10.1063/1.4826905}
	{\bibfield  {journal} {\bibinfo  {journal} {J. Appl. Phys.}\ }\textbf
		{\bibinfo {volume} {114}},\ \bibinfo {pages} {163508} (\bibinfo {year}
		{2013})}\BibitemShut {NoStop}%
	\bibitem [{\citenamefont {Yang}\ \emph {et~al.}(2021)\citenamefont {Yang},
		\citenamefont {Song}, \citenamefont {Lu}, \citenamefont {Zhang},
		\citenamefont {Zhang}, \citenamefont {Ni}, \citenamefont {Wang},
		\citenamefont {Li}, \citenamefont {Gu}, \citenamefont {Xie}, \citenamefont
		{Gao},\ and\ \citenamefont {Lou}}]{Yang2021}%
	\BibitemOpen
	\bibfield  {author} {\bibinfo {author} {\bibfnamefont {Y.}~\bibnamefont
			{Yang}}, \bibinfo {author} {\bibfnamefont {Z.}~\bibnamefont {Song}}, \bibinfo
		{author} {\bibfnamefont {G.}~\bibnamefont {Lu}}, \bibinfo {author}
		{\bibfnamefont {Q.}~\bibnamefont {Zhang}}, \bibinfo {author} {\bibfnamefont
			{B.}~\bibnamefont {Zhang}}, \bibinfo {author} {\bibfnamefont
			{B.}~\bibnamefont {Ni}}, \bibinfo {author} {\bibfnamefont {C.}~\bibnamefont
			{Wang}}, \bibinfo {author} {\bibfnamefont {X.}~\bibnamefont {Li}}, \bibinfo
		{author} {\bibfnamefont {L.}~\bibnamefont {Gu}}, \bibinfo {author}
		{\bibfnamefont {X.}~\bibnamefont {Xie}}, \bibinfo {author} {\bibfnamefont
			{H.}~\bibnamefont {Gao}},\ and\ \bibinfo {author} {\bibfnamefont
			{J.}~\bibnamefont {Lou}},\ }\bibfield  {title} {\bibinfo {title} {Intrinsic
			toughening and stable crack propagation in hexagonal boron nitride},\ }\href
	{https://doi.org/10.1038/s41586-021-03488-1} {\bibfield  {journal} {\bibinfo
			{journal} {Nature}\ }\textbf {\bibinfo {volume} {594}},\ \bibinfo {pages}
		{57} (\bibinfo {year} {2021})}\BibitemShut {NoStop}%
	\bibitem [{\citenamefont {Urade}\ \emph {et~al.}(2023)\citenamefont {Urade},
		\citenamefont {Lahiri},\ and\ \citenamefont {Suresh}}]{graphene_synthesis}%
	\BibitemOpen
	\bibfield  {author} {\bibinfo {author} {\bibfnamefont {A.~R.}\ \bibnamefont
			{Urade}}, \bibinfo {author} {\bibfnamefont {I.}~\bibnamefont {Lahiri}},\ and\
		\bibinfo {author} {\bibfnamefont {K.}~\bibnamefont {Suresh}},\ }\bibfield
	{title} {\bibinfo {title} {Graphene properties, synthesis and applications: a
			review},\ }\href@noop {} {\bibfield  {journal} {\bibinfo  {journal} {Jom}\
		}\textbf {\bibinfo {volume} {75}},\ \bibinfo {pages} {614} (\bibinfo {year}
		{2023})}\BibitemShut {NoStop}%
	\bibitem [{\citenamefont {Kashiwaya}\ \emph {et~al.}(2024)\citenamefont
		{Kashiwaya}, \citenamefont {Shi}, \citenamefont {Lu}, \citenamefont
		{Sangiovanni}, \citenamefont {Greczynski}, \citenamefont {Magnuson},
		\citenamefont {Andersson}, \citenamefont {Rosen},\ and\ \citenamefont
		{Hultman}}]{goldene_2024}%
	\BibitemOpen
	\bibfield  {author} {\bibinfo {author} {\bibfnamefont {S.}~\bibnamefont
			{Kashiwaya}}, \bibinfo {author} {\bibfnamefont {Y.}~\bibnamefont {Shi}},
		\bibinfo {author} {\bibfnamefont {J.}~\bibnamefont {Lu}}, \bibinfo {author}
		{\bibfnamefont {D.~G.}\ \bibnamefont {Sangiovanni}}, \bibinfo {author}
		{\bibfnamefont {G.}~\bibnamefont {Greczynski}}, \bibinfo {author}
		{\bibfnamefont {M.}~\bibnamefont {Magnuson}}, \bibinfo {author}
		{\bibfnamefont {M.}~\bibnamefont {Andersson}}, \bibinfo {author}
		{\bibfnamefont {J.}~\bibnamefont {Rosen}},\ and\ \bibinfo {author}
		{\bibfnamefont {L.}~\bibnamefont {Hultman}},\ }\bibfield  {title} {\bibinfo
		{title} {{Synthesis of goldene comprising single-atom layer gold}},\ }\href
	{https://doi.org/10.1038/s44160-024-00518-4} {\bibfield  {journal} {\bibinfo
			{journal} {Nat. Synth.}\ }\textbf {\bibinfo {volume} {3}},\ \bibinfo {pages}
		{744} (\bibinfo {year} {2024})}\BibitemShut {NoStop}%
	\bibitem [{\citenamefont {Naclerio}\ and\ \citenamefont
		{Kidambi}(2023)}]{hBN_synthesis}%
	\BibitemOpen
	\bibfield  {author} {\bibinfo {author} {\bibfnamefont {A.~E.}\ \bibnamefont
			{Naclerio}}\ and\ \bibinfo {author} {\bibfnamefont {P.~R.}\ \bibnamefont
			{Kidambi}},\ }\bibfield  {title} {\bibinfo {title} {A review of scalable
			hexagonal boron nitride (h-{BN}) synthesis for present and future
			applications},\ }\href
	{https://doi.org/https://doi.org/10.1002/adma.202207374} {\bibfield
		{journal} {\bibinfo  {journal} {Adv. Mater.}\ }\textbf {\bibinfo {volume}
			{35}},\ \bibinfo {pages} {2207374} (\bibinfo {year} {2023})}\BibitemShut
	{NoStop}%
	\bibitem [{\citenamefont {Cao}\ \emph {et~al.}(2015)\citenamefont {Cao},
		\citenamefont {Shen}, \citenamefont {Liang}, \citenamefont {Chen},\ and\
		\citenamefont {Shu}}]{Cao2015}%
	\BibitemOpen
	\bibfield  {author} {\bibinfo {author} {\bibfnamefont {D.}~\bibnamefont
			{Cao}}, \bibinfo {author} {\bibfnamefont {T.}~\bibnamefont {Shen}}, \bibinfo
		{author} {\bibfnamefont {P.}~\bibnamefont {Liang}}, \bibinfo {author}
		{\bibfnamefont {X.}~\bibnamefont {Chen}},\ and\ \bibinfo {author}
		{\bibfnamefont {H.}~\bibnamefont {Shu}},\ }\bibfield  {title} {\bibinfo
		{title} {Role of chemical potential in flake shape and edge properties of
			monolayer {MoS}$_2$},\ }\href {https://doi.org/10.1021/jp5097713} {\bibfield
		{journal} {\bibinfo  {journal} {J. Phys. Chem. C}\ }\textbf {\bibinfo
			{volume} {119}},\ \bibinfo {pages} {4294} (\bibinfo {year}
		{2015})}\BibitemShut {NoStop}%
	\bibitem [{\citenamefont {Elstner}\ \emph {et~al.}(1998)\citenamefont
		{Elstner}, \citenamefont {Porezag}, \citenamefont {Jungnickel}, \citenamefont
		{Elsner}, \citenamefont {Haugk}, \citenamefont {Frauenheim}, \citenamefont
		{Suhai},\ and\ \citenamefont {Seifert}}]{SCC}%
	\BibitemOpen
	\bibfield  {author} {\bibinfo {author} {\bibfnamefont {M.}~\bibnamefont
			{Elstner}}, \bibinfo {author} {\bibfnamefont {D.}~\bibnamefont {Porezag}},
		\bibinfo {author} {\bibfnamefont {G.}~\bibnamefont {Jungnickel}}, \bibinfo
		{author} {\bibfnamefont {J.}~\bibnamefont {Elsner}}, \bibinfo {author}
		{\bibfnamefont {M.}~\bibnamefont {Haugk}}, \bibinfo {author} {\bibfnamefont
			{T.}~\bibnamefont {Frauenheim}}, \bibinfo {author} {\bibfnamefont
			{S.}~\bibnamefont {Suhai}},\ and\ \bibinfo {author} {\bibfnamefont
			{G.}~\bibnamefont {Seifert}},\ }\bibfield  {title} {\bibinfo {title}
		{Self-consistent-charge density-functional tight-binding method for
			simulations of complex materials properties},\ }\href
	{https://doi.org/10.1103/PhysRevB.58.7260} {\bibfield  {journal} {\bibinfo
			{journal} {Phys. Rev. B}\ }\textbf {\bibinfo {volume} {58}},\ \bibinfo
		{pages} {7260} (\bibinfo {year} {1998})}\BibitemShut {NoStop}%
	\bibitem [{\citenamefont {Koskinen}(2012)}]{koskinen2012graphene}%
	\BibitemOpen
	\bibfield  {author} {\bibinfo {author} {\bibfnamefont {P.}~\bibnamefont
			{Koskinen}},\ }\bibfield  {title} {\bibinfo {title} {Graphene nanoribbons
			subject to gentle bends},\ }\href@noop {} {\bibfield  {journal} {\bibinfo
			{journal} {Phys. Rev. B}\ }\textbf {\bibinfo {volume} {85}},\ \bibinfo
		{pages} {205429} (\bibinfo {year} {2012})}\BibitemShut {NoStop}%
	\bibitem [{\citenamefont {Koskinen}\ \emph {et~al.}(2007)\citenamefont
		{Koskinen}, \citenamefont {H{\"a}kkinen}, \citenamefont {Huber},
		\citenamefont {von Issendorff},\ and\ \citenamefont
		{Moseler}}]{koskinen2007liquid}%
	\BibitemOpen
	\bibfield  {author} {\bibinfo {author} {\bibfnamefont {P.}~\bibnamefont
			{Koskinen}}, \bibinfo {author} {\bibfnamefont {H.}~\bibnamefont
			{H{\"a}kkinen}}, \bibinfo {author} {\bibfnamefont {B.}~\bibnamefont {Huber}},
		\bibinfo {author} {\bibfnamefont {B.}~\bibnamefont {von Issendorff}},\ and\
		\bibinfo {author} {\bibfnamefont {M.}~\bibnamefont {Moseler}},\ }\bibfield
	{title} {\bibinfo {title} {Liquid-liquid phase coexistence in gold clusters:
			{2D} or not {2D}?},\ }\href@noop {} {\bibfield  {journal} {\bibinfo
			{journal} {Phys. Rev. Lett.}\ }\textbf {\bibinfo {volume} {98}},\ \bibinfo
		{pages} {015701} (\bibinfo {year} {2007})}\BibitemShut {NoStop}%
	\bibitem [{\citenamefont {Koskinen}\ and\ \citenamefont
		{Korhonen}(2015)}]{koskinen2015plenty}%
	\BibitemOpen
	\bibfield  {author} {\bibinfo {author} {\bibfnamefont {P.}~\bibnamefont
			{Koskinen}}\ and\ \bibinfo {author} {\bibfnamefont {T.}~\bibnamefont
			{Korhonen}},\ }\bibfield  {title} {\bibinfo {title} {Plenty of motion at the
			bottom: Atomically thin liquid gold membrane},\ }\href@noop {} {\bibfield
		{journal} {\bibinfo  {journal} {Nanoscale}\ }\textbf {\bibinfo {volume}
			{7}},\ \bibinfo {pages} {10140} (\bibinfo {year} {2015})}\BibitemShut
	{NoStop}%
	\bibitem [{\citenamefont {Koskinen}\ \emph {et~al.}(2014)\citenamefont
		{Koskinen}, \citenamefont {Fampiou},\ and\ \citenamefont
		{Ramasubramaniam}}]{koskinen2014density}%
	\BibitemOpen
	\bibfield  {author} {\bibinfo {author} {\bibfnamefont {P.}~\bibnamefont
			{Koskinen}}, \bibinfo {author} {\bibfnamefont {I.}~\bibnamefont {Fampiou}},\
		and\ \bibinfo {author} {\bibfnamefont {A.}~\bibnamefont {Ramasubramaniam}},\
	}\bibfield  {title} {\bibinfo {title} {Density-functional tight-binding
			simulations of curvature-controlled layer decoupling and band-gap tuning in
			bilayer {MoS}$_2$},\ }\href@noop {} {\bibfield  {journal} {\bibinfo
			{journal} {Phys. Rev. Lett.}\ }\textbf {\bibinfo {volume} {112}},\ \bibinfo
		{pages} {186802} (\bibinfo {year} {2014})}\BibitemShut {NoStop}%
	\bibitem [{\citenamefont {Zobelli}\ \emph {et~al.}(2007)\citenamefont
		{Zobelli}, \citenamefont {Ewels}, \citenamefont {Gloter},\ and\ \citenamefont
		{Seifert}}]{zobelli2007vacancy}%
	\BibitemOpen
	\bibfield  {author} {\bibinfo {author} {\bibfnamefont {A.}~\bibnamefont
			{Zobelli}}, \bibinfo {author} {\bibfnamefont {C.}~\bibnamefont {Ewels}},
		\bibinfo {author} {\bibfnamefont {A.}~\bibnamefont {Gloter}},\ and\ \bibinfo
		{author} {\bibfnamefont {G.}~\bibnamefont {Seifert}},\ }\bibfield  {title}
	{\bibinfo {title} {Vacancy migration in hexagonal boron nitride},\
	}\href@noop {} {\bibfield  {journal} {\bibinfo  {journal} {Phys. Rev. B}\
		}\textbf {\bibinfo {volume} {75}},\ \bibinfo {pages} {094104} (\bibinfo
		{year} {2007})}\BibitemShut {NoStop}%
	\bibitem [{\citenamefont {Koskinen}\ and\ \citenamefont
		{Mäkinen}(2009)}]{hotbit}%
	\BibitemOpen
	\bibfield  {author} {\bibinfo {author} {\bibfnamefont {P.}~\bibnamefont
			{Koskinen}}\ and\ \bibinfo {author} {\bibfnamefont {V.}~\bibnamefont
			{Mäkinen}},\ }\bibfield  {title} {\bibinfo {title} {Density-functional
			tight-binding for beginners},\ }\href@noop {} {\bibfield  {journal} {\bibinfo
			{journal} {Comp. Mat. Sci.}\ }\textbf {\bibinfo {volume} {47}},\ \bibinfo
		{pages} {237} (\bibinfo {year} {2009})}\BibitemShut {NoStop}%
	\bibitem [{\citenamefont {Harrison}\ \emph {et~al.}(2018)\citenamefont
		{Harrison}, \citenamefont {Schall}, \citenamefont {Maskey}, \citenamefont
		{Mikulski}, \citenamefont {Knippenberg},\ and\ \citenamefont {Morrow}}]{ff}%
	\BibitemOpen
	\bibfield  {author} {\bibinfo {author} {\bibfnamefont {J.~A.}\ \bibnamefont
			{Harrison}}, \bibinfo {author} {\bibfnamefont {J.~D.}\ \bibnamefont
			{Schall}}, \bibinfo {author} {\bibfnamefont {S.}~\bibnamefont {Maskey}},
		\bibinfo {author} {\bibfnamefont {P.~T.}\ \bibnamefont {Mikulski}}, \bibinfo
		{author} {\bibfnamefont {M.~T.}\ \bibnamefont {Knippenberg}},\ and\ \bibinfo
		{author} {\bibfnamefont {B.~H.}\ \bibnamefont {Morrow}},\ }\bibfield  {title}
	{\bibinfo {title} {Review of force fields and intermolecular potentials used
			in atomistic computational materials research},\ }\href
	{https://doi.org/10.1063/1.5020808} {\bibfield  {journal} {\bibinfo
			{journal} {Appl. Phys. Rev.}\ }\textbf {\bibinfo {volume} {5}},\ \bibinfo
		{pages} {031104} (\bibinfo {year} {2018})}\BibitemShut {NoStop}%
	\bibitem [{\citenamefont {Pinheiro}\ \emph {et~al.}(2021)\citenamefont
		{Pinheiro}, \citenamefont {Ge}, \citenamefont {Ferre}, \citenamefont
		{Dral},\ and\ \citenamefont {Barbatti}}]{ml}%
	\BibitemOpen
	\bibfield  {author} {\bibinfo {author} {\bibfnamefont {M.}~\bibnamefont
			{Pinheiro}}, \bibinfo {author} {\bibfnamefont {F.}~\bibnamefont {Ge}},
		\bibinfo {author} {\bibfnamefont {N.}~\bibnamefont {Ferr{\'e}}}, \bibinfo
		{author} {\bibfnamefont {P.~O.}\ \bibnamefont {Dral}},\ and\ \bibinfo
		{author} {\bibfnamefont {M.}~\bibnamefont {Barbatti}},\ }\bibfield  {title}
	{\bibinfo {title} {Choosing the right molecular machine learning potential},\
	}\href@noop {} {\bibfield  {journal} {\bibinfo  {journal} {Chem. Sci.}\
		}\textbf {\bibinfo {volume} {12}},\ \bibinfo {pages} {14396} (\bibinfo {year}
		{2021})}\BibitemShut {NoStop}%
	\bibitem [{\citenamefont {Frenzel}\ \emph {et~al.}(2004)\citenamefont
		{Frenzel}, \citenamefont {Oliveira}, \citenamefont {Jardillier},
		\citenamefont {Heine},\ and\ \citenamefont {Seifert}}]{matsci}%
	\BibitemOpen
	\bibfield  {author} {\bibinfo {author} {\bibfnamefont {J.}~\bibnamefont
			{Frenzel}}, \bibinfo {author} {\bibfnamefont {A.}~\bibnamefont {Oliveira}},
		\bibinfo {author} {\bibfnamefont {N.}~\bibnamefont {Jardillier}}, \bibinfo
		{author} {\bibfnamefont {T.}~\bibnamefont {Heine}},\ and\ \bibinfo {author}
		{\bibfnamefont {G.}~\bibnamefont {Seifert}},\ }\bibfield  {title} {\bibinfo
		{title} {Semi-relativistic, self-consistent charge slater-koster tables for
			density-functional based tight-binding ({DFTB}) for materials science
			simulations},\ }\href@noop {} {\bibfield  {journal} {\bibinfo  {journal}
			{Zeolites}\ }\textbf {\bibinfo {volume} {2}},\ \bibinfo {pages} {7} (\bibinfo
		{year} {2004})}\BibitemShut {NoStop}%
	\bibitem [{\citenamefont {Yue}\ \emph {et~al.}(2017)\citenamefont {Yue},
		\citenamefont {Seifert}, \citenamefont {Chang},\ and\ \citenamefont
		{Zhang}}]{hBN_dftb}%
	\BibitemOpen
	\bibfield  {author} {\bibinfo {author} {\bibfnamefont {L.}~\bibnamefont
			{Yue}}, \bibinfo {author} {\bibfnamefont {G.}~\bibnamefont {Seifert}},
		\bibinfo {author} {\bibfnamefont {K.}~\bibnamefont {Chang}},\ and\ \bibinfo
		{author} {\bibfnamefont {D.-B.}\ \bibnamefont {Zhang}},\ }\bibfield  {title}
	{\bibinfo {title} {Effective zeeman splitting in bent lateral heterojunctions
			of graphene and hexagonal boron nitride: A new mechanism towards
			half-metallicity},\ }\href {https://doi.org/10.1103/PhysRevB.96.201403}
	{\bibfield  {journal} {\bibinfo  {journal} {Phys. Rev. B}\ }\textbf {\bibinfo
			{volume} {96}},\ \bibinfo {pages} {201403} (\bibinfo {year}
		{2017})}\BibitemShut {NoStop}%
	\bibitem [{\citenamefont {Porezag}\ \emph {et~al.}(1995)\citenamefont
		{Porezag}, \citenamefont {Frauenheim}, \citenamefont {K\"ohler},
		\citenamefont {Seifert},\ and\ \citenamefont {Kaschner}}]{hotbit_param_old}%
	\BibitemOpen
	\bibfield  {author} {\bibinfo {author} {\bibfnamefont {D.}~\bibnamefont
			{Porezag}}, \bibinfo {author} {\bibfnamefont {T.}~\bibnamefont {Frauenheim}},
		\bibinfo {author} {\bibfnamefont {T.}~\bibnamefont {K\"ohler}}, \bibinfo
		{author} {\bibfnamefont {G.}~\bibnamefont {Seifert}},\ and\ \bibinfo {author}
		{\bibfnamefont {R.}~\bibnamefont {Kaschner}},\ }\bibfield  {title} {\bibinfo
		{title} {Construction of tight-binding-like potentials on the basis of
			density-functional theory: Application to carbon},\ }\href
	{https://doi.org/10.1103/PhysRevB.51.12947} {\bibfield  {journal} {\bibinfo
			{journal} {Phys. Rev. B}\ }\textbf {\bibinfo {volume} {51}},\ \bibinfo
		{pages} {12947} (\bibinfo {year} {1995})}\BibitemShut {NoStop}%
	\bibitem [{\citenamefont {Seifert}\ \emph {et~al.}(2000)\citenamefont
		{Seifert}, \citenamefont {Terrones}, \citenamefont {Terrones}, \citenamefont
		{Jungnickel},\ and\ \citenamefont {Frauenheim}}]{hotbit_param_mos2}%
	\BibitemOpen
	\bibfield  {author} {\bibinfo {author} {\bibfnamefont {G.}~\bibnamefont
			{Seifert}}, \bibinfo {author} {\bibfnamefont {H.}~\bibnamefont {Terrones}},
		\bibinfo {author} {\bibfnamefont {M.}~\bibnamefont {Terrones}}, \bibinfo
		{author} {\bibfnamefont {G.}~\bibnamefont {Jungnickel}},\ and\ \bibinfo
		{author} {\bibfnamefont {T.}~\bibnamefont {Frauenheim}},\ }\bibfield  {title}
	{\bibinfo {title} {Structure and electronic properties of
			${\mathrm{mos}}_{2}$ nanotubes},\ }\href
	{https://doi.org/10.1103/PhysRevLett.85.146} {\bibfield  {journal} {\bibinfo
			{journal} {Phys. Rev. Lett.}\ }\textbf {\bibinfo {volume} {85}},\ \bibinfo
		{pages} {146149} (\bibinfo {year} {2000})}\BibitemShut {NoStop}%
	\bibitem [{\citenamefont {Mäkinen}\ \emph {et~al.}(2013)\citenamefont
		{Mäkinen}, \citenamefont {Koskinen},\ and\ \citenamefont
		{Häkkinen}}]{hotbit_param_gold}%
	\BibitemOpen
	\bibfield  {author} {\bibinfo {author} {\bibfnamefont {V.}~\bibnamefont
			{Mäkinen}}, \bibinfo {author} {\bibfnamefont {P.}~\bibnamefont {Koskinen}},\
		and\ \bibinfo {author} {\bibfnamefont {H.}~\bibnamefont {Häkkinen}},\
	}\bibfield  {title} {\bibinfo {title} {Modeling thiolate-protected gold
			clusters with density-functional tight-binding},\ }\href
	{https://doi.org/10.1140/epjd/e2012-30486-4} {\bibfield  {journal} {\bibinfo
			{journal} {Eur. Phys. J. D}\ }\textbf {\bibinfo {volume} {67}},\ \bibinfo
		{pages} {14346079} (\bibinfo {year} {2013})}\BibitemShut {NoStop}%
	\bibitem [{\citenamefont {Nocedal}\ and\ \citenamefont
		{Wright}(1999)}]{nocedal1999numerical}%
	\BibitemOpen
	\bibfield  {author} {\bibinfo {author} {\bibfnamefont {J.}~\bibnamefont
			{Nocedal}}\ and\ \bibinfo {author} {\bibfnamefont {S.~J.}\ \bibnamefont
			{Wright}},\ }\href@noop {} {\emph {\bibinfo {title} {Numerical
				optimization}}}\ (\bibinfo  {publisher} {Springer},\ \bibinfo {year}
	{1999})\BibitemShut {NoStop}%
	\bibitem [{\citenamefont {Methfessel}\ \emph {et~al.}(1992)\citenamefont
		{Methfessel}, \citenamefont {Hennig},\ and\ \citenamefont
		{Scheffler}}]{bondCutting1992calculated}%
	\BibitemOpen
	\bibfield  {author} {\bibinfo {author} {\bibfnamefont {M.}~\bibnamefont
			{Methfessel}}, \bibinfo {author} {\bibfnamefont {D.}~\bibnamefont {Hennig}},\
		and\ \bibinfo {author} {\bibfnamefont {M.}~\bibnamefont {Scheffler}},\
	}\bibfield  {title} {\bibinfo {title} {Calculated surface energies of the 4d
			transition metals: A study of bond-cutting models},\ }\href@noop {}
	{\bibfield  {journal} {\bibinfo  {journal} {Appl. Phys. A}\ }\textbf
		{\bibinfo {volume} {55}},\ \bibinfo {pages} {442} (\bibinfo {year}
		{1992})}\BibitemShut {NoStop}%
	\bibitem [{\citenamefont {Ruvireta}\ \emph {et~al.}(2017)\citenamefont
		{Ruvireta}, \citenamefont {Vega},\ and\ \citenamefont
		{Viñes}}]{CohCoord2017}%
	\BibitemOpen
	\bibfield  {author} {\bibinfo {author} {\bibfnamefont {J.}~\bibnamefont
			{Ruvireta}}, \bibinfo {author} {\bibfnamefont {L.}~\bibnamefont {Vega}},\
		and\ \bibinfo {author} {\bibfnamefont {F.}~\bibnamefont {Viñes}},\
	}\bibfield  {title} {\bibinfo {title} {Cohesion and coordination effects on
			transition metal surface energies},\ }\href
	{https://doi.org/https://doi.org/10.1016/j.susc.2017.05.013} {\bibfield
		{journal} {\bibinfo  {journal} {Surface Science}\ }\textbf {\bibinfo {volume}
			{664}},\ \bibinfo {pages} {45} (\bibinfo {year} {2017})}\BibitemShut
	{NoStop}%
	\bibitem [{\citenamefont
		{McLachlan~Jr}(1957)}]{Bond-cuttingmclachlan1957surface}%
	\BibitemOpen
	\bibfield  {author} {\bibinfo {author} {\bibfnamefont {D.}~\bibnamefont
			{McLachlan~Jr}},\ }\bibfield  {title} {\bibinfo {title} {The surface tension
			of solid metals},\ }\href@noop {} {\bibfield  {journal} {\bibinfo  {journal}
			{Acta Metall.}\ }\textbf {\bibinfo {volume} {5}},\ \bibinfo {pages} {111}
		(\bibinfo {year} {1957})}\BibitemShut {NoStop}%
	\bibitem [{\citenamefont {Lee}\ \emph {et~al.}(2008)\citenamefont {Lee},
		\citenamefont {Wei}, \citenamefont {Kysar},\ and\ \citenamefont
		{Hone}}]{Changgu2008}%
	\BibitemOpen
	\bibfield  {author} {\bibinfo {author} {\bibfnamefont {C.}~\bibnamefont
			{Lee}}, \bibinfo {author} {\bibfnamefont {X.}~\bibnamefont {Wei}}, \bibinfo
		{author} {\bibfnamefont {J.~W.}\ \bibnamefont {Kysar}},\ and\ \bibinfo
		{author} {\bibfnamefont {J.}~\bibnamefont {Hone}},\ }\bibfield  {title}
	{\bibinfo {title} {Measurement of the elastic properties and intrinsic
			strength of monolayer graphene},\ }\href
	{https://doi.org/10.1126/science.1157996} {\bibfield  {journal} {\bibinfo
			{journal} {Science}\ }\textbf {\bibinfo {volume} {321}},\ \bibinfo {pages}
		{385} (\bibinfo {year} {2008})}\BibitemShut {NoStop}%
	\bibitem [{\citenamefont {Falin}\ \emph {et~al.}(2017)\citenamefont {Falin},
		\citenamefont {Cai}, \citenamefont {Santos}, \citenamefont {Scullion},
		\citenamefont {Qian}, \citenamefont {Zhang}, \citenamefont {Yang},
		\citenamefont {Huang}, \citenamefont {Watanabe}, \citenamefont {Taniguchi},
		\citenamefont {Barnett}, \citenamefont {Chen}, \citenamefont {Ruoff},\ and\
		\citenamefont {Li}}]{Falin2017}%
	\BibitemOpen
	\bibfield  {author} {\bibinfo {author} {\bibfnamefont {A.}~\bibnamefont
			{Falin}}, \bibinfo {author} {\bibfnamefont {Q.}~\bibnamefont {Cai}}, \bibinfo
		{author} {\bibfnamefont {E.~J.}\ \bibnamefont {Santos}}, \bibinfo {author}
		{\bibfnamefont {D.}~\bibnamefont {Scullion}}, \bibinfo {author}
		{\bibfnamefont {D.}~\bibnamefont {Qian}}, \bibinfo {author} {\bibfnamefont
			{R.}~\bibnamefont {Zhang}}, \bibinfo {author} {\bibfnamefont
			{Z.}~\bibnamefont {Yang}}, \bibinfo {author} {\bibfnamefont {S.}~\bibnamefont
			{Huang}}, \bibinfo {author} {\bibfnamefont {K.}~\bibnamefont {Watanabe}},
		\bibinfo {author} {\bibfnamefont {T.}~\bibnamefont {Taniguchi}}, \bibinfo
		{author} {\bibfnamefont {M.~R.}\ \bibnamefont {Barnett}}, \bibinfo {author}
		{\bibfnamefont {Y.}~\bibnamefont {Chen}}, \bibinfo {author} {\bibfnamefont
			{R.~S.}\ \bibnamefont {Ruoff}},\ and\ \bibinfo {author} {\bibfnamefont
			{L.~H.}\ \bibnamefont {Li}},\ }\bibfield  {title} {\bibinfo {title}
		{Mechanical properties of atomically thin boron nitride and the role of
			interlayer interactions},\ }\href {https://doi.org/10.1038/ncomms15815}
	{\bibfield  {journal} {\bibinfo  {journal} {Nat. Commun.}\ }\textbf {\bibinfo
			{volume} {8}},\ \bibinfo {pages} {15815} (\bibinfo {year}
		{2017})}\BibitemShut {NoStop}%
	\bibitem [{\citenamefont {Mortazavi}(2024)}]{Mortazavi2024}%
	\BibitemOpen
	\bibfield  {author} {\bibinfo {author} {\bibfnamefont {B.}~\bibnamefont
			{Mortazavi}},\ }\bibfield  {title} {\bibinfo {title} {Goldene: An anisotropic
			metallic monolayer with remarkable stability and rigidity and low lattice
			thermal conductivity},\ }\href {https://www.mdpi.com/1996-1944/17/11/2653}
	{\bibfield  {journal} {\bibinfo  {journal} {Materials}\ }\textbf {\bibinfo
			{volume} {17}},\ \bibinfo {pages} {2653} (\bibinfo {year}
		{2024})}\BibitemShut {NoStop}%
	\bibitem [{\citenamefont {Bertolazzi}\ \emph {et~al.}(2011)\citenamefont
		{Bertolazzi}, \citenamefont {Brivio},\ and\ \citenamefont
		{Kis}}]{Bertolazzi2011}%
	\BibitemOpen
	\bibfield  {author} {\bibinfo {author} {\bibfnamefont {S.}~\bibnamefont
			{Bertolazzi}}, \bibinfo {author} {\bibfnamefont {J.}~\bibnamefont {Brivio}},\
		and\ \bibinfo {author} {\bibfnamefont {A.}~\bibnamefont {Kis}},\ }\bibfield
	{title} {\bibinfo {title} {Stretching and breaking of ultrathin {MoS}$_2$},\
	}\href {https://doi.org/10.1021/nn203879f} {\bibfield  {journal} {\bibinfo
			{journal} {ACS Nano}\ }\textbf {\bibinfo {volume} {5}},\ \bibinfo {pages}
		{9703} (\bibinfo {year} {2011})}\BibitemShut {NoStop}%
	\bibitem [{\citenamefont {Koskinen}\ \emph {et~al.}(2008)\citenamefont
		{Koskinen}, \citenamefont {Malola},\ and\ \citenamefont
		{H{\"a}kkinen}}]{koskinen2008self}%
	\BibitemOpen
	\bibfield  {author} {\bibinfo {author} {\bibfnamefont {P.}~\bibnamefont
			{Koskinen}}, \bibinfo {author} {\bibfnamefont {S.}~\bibnamefont {Malola}},\
		and\ \bibinfo {author} {\bibfnamefont {H.}~\bibnamefont {H{\"a}kkinen}},\
	}\bibfield  {title} {\bibinfo {title} {Self-passivating edge reconstructions
			of graphene},\ }\href@noop {} {\bibfield  {journal} {\bibinfo  {journal}
			{Physical review letters}\ }\textbf {\bibinfo {volume} {101}},\ \bibinfo
		{pages} {115502} (\bibinfo {year} {2008})}\BibitemShut {NoStop}%
	\bibitem [{\citenamefont {Basu}\ \emph {et~al.}(2008)\citenamefont {Basu},
		\citenamefont {Gilbert}, \citenamefont {Register}, \citenamefont {Banerjee},\
		and\ \citenamefont {MacDonald}}]{basu2008effect}%
	\BibitemOpen
	\bibfield  {author} {\bibinfo {author} {\bibfnamefont {D.}~\bibnamefont
			{Basu}}, \bibinfo {author} {\bibfnamefont {M.}~\bibnamefont {Gilbert}},
		\bibinfo {author} {\bibfnamefont {L.}~\bibnamefont {Register}}, \bibinfo
		{author} {\bibfnamefont {S.~K.}\ \bibnamefont {Banerjee}},\ and\ \bibinfo
		{author} {\bibfnamefont {A.~H.}\ \bibnamefont {MacDonald}},\ }\bibfield
	{title} {\bibinfo {title} {Effect of edge roughness on electronic transport
			in graphene nanoribbon channel metal-oxide-semiconductor field-effect
			transistors},\ }\href@noop {} {\bibfield  {journal} {\bibinfo  {journal}
			{Appl. Phys. Lett.}\ }\textbf {\bibinfo {volume} {92}} (\bibinfo {year}
		{2008})}\BibitemShut {NoStop}%
	\bibitem [{\citenamefont {Girit}\ \emph {et~al.}(2009)\citenamefont {Girit},
		\citenamefont {Meyer}, \citenamefont {Erni}, \citenamefont {Rossell},
		\citenamefont {Kisielowski}, \citenamefont {Yang}, \citenamefont {Park},
		\citenamefont {Crommie}, \citenamefont {Cohen}, \citenamefont {Louie} \emph
		{et~al.}}]{girit2009graphene}%
	\BibitemOpen
	\bibfield  {author} {\bibinfo {author} {\bibfnamefont {C.~O.}\ \bibnamefont
			{Girit}}, \bibinfo {author} {\bibfnamefont {J.~C.}\ \bibnamefont {Meyer}},
		\bibinfo {author} {\bibfnamefont {R.}~\bibnamefont {Erni}}, \bibinfo {author}
		{\bibfnamefont {M.~D.}\ \bibnamefont {Rossell}}, \bibinfo {author}
		{\bibfnamefont {C.}~\bibnamefont {Kisielowski}}, \bibinfo {author}
		{\bibfnamefont {L.}~\bibnamefont {Yang}}, \bibinfo {author} {\bibfnamefont
			{C.-H.}\ \bibnamefont {Park}}, \bibinfo {author} {\bibfnamefont
			{M.}~\bibnamefont {Crommie}}, \bibinfo {author} {\bibfnamefont {M.~L.}\
			\bibnamefont {Cohen}}, \bibinfo {author} {\bibfnamefont {S.~G.}\ \bibnamefont
			{Louie}}, \emph {et~al.},\ }\bibfield  {title} {\bibinfo {title} {Graphene at
			the edge: stability and dynamics},\ }\href@noop {} {\bibfield  {journal}
		{\bibinfo  {journal} {Science}\ }\textbf {\bibinfo {volume} {323}},\ \bibinfo
		{pages} {1705} (\bibinfo {year} {2009})}\BibitemShut {NoStop}%
	\bibitem [{\citenamefont {Yin}\ \emph {et~al.}(2009)\citenamefont {Yin},
		\citenamefont {Akola}, \citenamefont {Koskinen}, \citenamefont {Manninen},\
		and\ \citenamefont {Palmer}}]{yin2009bright}%
	\BibitemOpen
	\bibfield  {author} {\bibinfo {author} {\bibfnamefont {F.}~\bibnamefont
			{Yin}}, \bibinfo {author} {\bibfnamefont {J.}~\bibnamefont {Akola}}, \bibinfo
		{author} {\bibfnamefont {P.}~\bibnamefont {Koskinen}}, \bibinfo {author}
		{\bibfnamefont {M.}~\bibnamefont {Manninen}},\ and\ \bibinfo {author}
		{\bibfnamefont {R.}~\bibnamefont {Palmer}},\ }\bibfield  {title} {\bibinfo
		{title} {Bright beaches of nanoscale potassium islands on graphite in {STM}
			imaging},\ }\href@noop {} {\bibfield  {journal} {\bibinfo  {journal} {Phys.
				Rev. Lett.}\ }\textbf {\bibinfo {volume} {102}},\ \bibinfo {pages} {106102}
		(\bibinfo {year} {2009})}\BibitemShut {NoStop}%
	\bibitem [{\citenamefont {Shenoy}\ \emph {et~al.}(2010)\citenamefont {Shenoy},
		\citenamefont {Reddy},\ and\ \citenamefont {Zhang}}]{shenoy2010spontaneous}%
	\BibitemOpen
	\bibfield  {author} {\bibinfo {author} {\bibfnamefont {V.~B.}\ \bibnamefont
			{Shenoy}}, \bibinfo {author} {\bibfnamefont {C.~D.}\ \bibnamefont {Reddy}},\
		and\ \bibinfo {author} {\bibfnamefont {Y.-W.}\ \bibnamefont {Zhang}},\
	}\bibfield  {title} {\bibinfo {title} {Spontaneous curling of graphene sheets
			with reconstructed edges},\ }\href@noop {} {\bibfield  {journal} {\bibinfo
			{journal} {ACS Nano}\ }\textbf {\bibinfo {volume} {4}},\ \bibinfo {pages}
		{4840} (\bibinfo {year} {2010})}\BibitemShut {NoStop}%
	\bibitem [{\citenamefont {Ramasubramaniam}\ \emph {et~al.}(2012)\citenamefont
		{Ramasubramaniam}, \citenamefont {Koskinen}, \citenamefont {Kit},\ and\
		\citenamefont {Shenoy}}]{EdgeStressInducedSpontaneous20212}%
	\BibitemOpen
	\bibfield  {author} {\bibinfo {author} {\bibfnamefont {A.}~\bibnamefont
			{Ramasubramaniam}}, \bibinfo {author} {\bibfnamefont {P.}~\bibnamefont
			{Koskinen}}, \bibinfo {author} {\bibfnamefont {O.~O.}\ \bibnamefont {Kit}},\
		and\ \bibinfo {author} {\bibfnamefont {V.~B.}\ \bibnamefont {Shenoy}},\
	}\bibfield  {title} {\bibinfo {title} {{Edge-stress-induced spontaneous
				twisting of graphene nanoribbons}},\ }\href
	{https://doi.org/10.1063/1.3689814} {\bibfield  {journal} {\bibinfo
			{journal} {J. Appl. Phys.}\ }\textbf {\bibinfo {volume} {111}},\ \bibinfo
		{pages} {054302} (\bibinfo {year} {2012})}\BibitemShut {NoStop}%
\end{thebibliography}
%

\end{document}